\documentclass[
  reprint,              % two-column journal layout
  aps,                  % APS styling (numeric superscript citations)
  prmaterials,          % pick an APS journal style; change if desired (e.g., prb, prl, prapplied)
  superscriptaddress,   % affiliations as superscripts
  longbibliography,     % show full titles/etc. in refs (optional)
  floatfix              % helps LaTeX place floats
]{revtex4-2}

\usepackage[version=4]{mhchem}
\usepackage{graphicx}% Include figure files
\usepackage{dcolumn}% Align table columns on decimal point
\usepackage{bm}% bold math
\usepackage{amsmath}
\usepackage{array}
\usepackage[breaklinks=true]{hyperref}
\usepackage{xurl}

\graphicspath{{./}{../figures/}}

\begin{document}

\preprint{APS/123-QED}

% ---- Title/author setup (RevTeX-native) ----
\title{Ferrocene-functionalized covalent organic framework exceeding the ultimate hydrogen storage targets: a first-principles multiscale computational study}

\author{Marcus Djokic}
\affiliation{Department of Chemical Engineering and Material Science, Michigan State University, East Lansing, MI 48824, USA}

\author{Jose L. Mendoza-Cortes}
\email{jmendoza@msu.edu}
\affiliation{Department of Chemical Engineering and Material Science, Michigan State University, East Lansing, MI 48824, USA}
\affiliation{Department of Physics and Astronomy, Michigan State University, East Lansing, MI 48824, USA}

%\date{} % suppress date

\begin{abstract}
The development of efficient hydrogen storage materials is crucial for advancing the hydrogen economy and meeting the U.S. Department of Energy's targets of 6.5 wt\% and 50 g \ce{H2} L$^{-1}$ for automotive applications. We present a computational study of ferrocene-functionalized covalent organic frameworks (COFs) for hydrogen storage. Following the \textbf{M}ulti-binding \textbf{S}ites \textbf{U}nited in \textbf{C}ovalent-\textbf{O}rganic \textbf{F}ramework (MSUCOF) approach, we introduce MSUCOF-4-FeCp, designed by incorporating ferrocene (\ce{FeCp2}) moieties into IRCOF-102. Notably, it achieves exceptional performance with gravimetric and volumetric uptakes of 18.0 wt\% and 72.6 g \ce{H2} L$^{-1}$ at 298 K and 700 bar. The material exhibits optimal binding energies (15--20 kJ$\cdot$mol$^{-1}$) ensuring both high storage capacity and deliverable hydrogen under practical conditions. This work establishes ferrocene functionalization as a cost-effective alternative to precious metal incorporation in COFs.
\end{abstract}

\maketitle

%%%MAIN TEXT%%%%
\section{Introduction}

The urgent need for sustainable energy solutions positions hydrogen as a promising clean energy carrier, capable of addressing both energy security and environmental concerns associated with the dependence on fossil fuels. However, the widespread adoption of hydrogen as a transportation fuel faces significant challenges, particularly in developing safe, efficient, and cost-effective storage systems. The U.S. Department of Energy (DOE) has established ambitious targets for hydrogen storage in light-duty vehicles: 6.5 wt\% gravimetric capacity and 50 g \ce{H2} L$^{-1}$ volumetric capacity, with the overarching goal of reducing clean hydrogen costs to \$1 per kilogram by 2031.\cite{doe_hydrogen_shot_2021,doe_hydrogen_targets_2020}

\subsection{Covalent organic frameworks for hydrogen storage}

Covalent organic frameworks (COFs) have emerged as particularly promising candidates for hydrogen storage applications due to their high surface areas, tunable pore structures, and composition primarily of light elements.\cite{Surf_Area} Unlike their metal-organic framework (MOF) counterparts, COFs offer the advantage of lower framework density while maintaining structural integrity and thermal stability. The crystalline nature of COFs allows for precise control over the size and functionality of pores, allowing rational design approaches to optimize their hydrogen uptake properties.\cite{wu_adsorption_2010,mendoza-cortes_adsorption_2010}

Recent advances in COF chemistry have demonstrated that hydrogen storage performance can be significantly enhanced through strategic incorporation of functional groups and heteroatoms.\cite{fan_multivariate_2021,humby_hostguest_2019} The introduction of open metal sites and polarizing groups has been shown to strengthen van der Waals interactions with hydrogen molecules, thus improving binding enthalpies and storage capacities.\cite{al-rowaili_review_2021,han_recent_2009} However, achieving the optimal balance between gravimetric and volumetric performance remains a central challenge, as strategies to increase volumetric uptake often lead to increased framework density and reduced gravimetric capacity.\cite{caskey_dramatic_2008}

\begin{figure*}[!htbp]
  \centering
  \includegraphics[width=\textwidth]{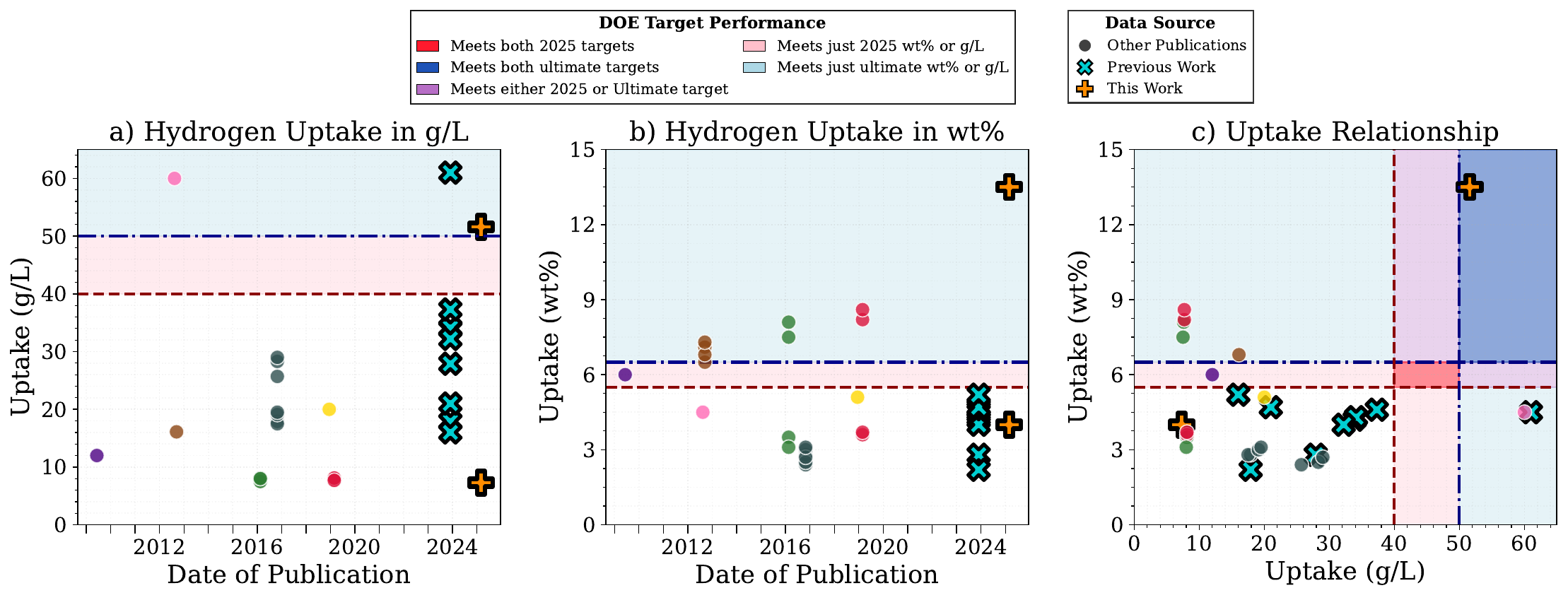}
  \caption{Leading MOF/COF hydrogen storage performance\cite{Klontzas, Guo, mendoza-cortes_covalent_2012, pramudya_design_2016, Xiao-Dong_2016, Xiao-Dong_2019, Ghosh, Djokic2024} compared to U.S.\ Department of Energy (DOE) light-duty vehicle (LDV) hydrogen storage targets.\cite{doe_hydrogen_targets_2020} Total hydrogen uptakes reported at 298~K and up to 100~bar are shown as: (a) volumetric uptake (g\,L$^{-1}$) versus publication year, (b) gravimetric uptake (wt\%) versus publication year, and (c) volumetric versus gravimetric uptake, illustrating the trade-off between storage density metrics. In (c), the DOE Ultimate LDV target corresponds to the upper-right quadrant (dark blue). Circles ($\circ$) represent literature-reported materials, $\times$ denotes our group’s prior MSUCOF study, and $+$ indicates results from the present work; identical colors indicate data points originating from the same publication.}
  \label{fig:LitReview}
\end{figure*}

\subsection{Metallocene chemistry and hydrogen binding}

Metallocenes represent a unique class of organometallic compounds characterized by their distinctive "sandwich" structure, where a central metal atom is coordinated to two cyclic organic ligands. The discovery of ferrocene in 1951 by Kealy and Pauson\cite{Kealy1951} revolutionized organometallic chemistry and earned its structural elucidators, Geoffrey Wilkinson\cite{Wilkinson1952} and Ernst Otto Fischer\cite{Fischer1952}, the 1973 Nobel Prize in Chemistry.\cite{Laszlo2000} Ferrocene (\ce{Fe(\eta^5-C5H5)2}) exhibits remarkable thermal stability (stable to $400^\circ \text{C}$), chemical inertness, and well-defined electrochemical properties with a reversible \ce{Fe^{2+}/Fe^{3+}} redox couple at +0.4 V \textit{vs.} SCE.\cite{Roy2022}

The electronic structure of ferrocene, governed by the 18-electron rule, provides multiple potential binding sites for small molecules; these binding sites have been exploited across various porous materials. Ferrocene-functionalized MOFs have already been synthesized, utilizing the redox-active iron center to enable photocatalytic reduction of \ce{CO2} with enhanced electron transfer capabilities.\cite{Lai2023} Similarly, microporous polymers containing ferrocene demonstrate improved storage of \ce{CO2} through the synergistic effects of microporosity and metal-ligand interactions.\cite{Samy2022} Analogously, superior hydrogen storage performance has also been reported for carbon foams loaded with ferrocene.\cite{Chen2012}

The versatility of ferrocene extends beyond simple gas adsorption: it fundamentally alters electronic density distributions, spatial arrangements, and coordination environments within composite materials, thereby modulating catalytic activity for HER/OER reactions.\cite{Sariga2023} In hypercrosslinked polymers, ferrocene incorporation achieves enhanced \ce{H2} adsorption through a dual mechanism; the abundant micropores provide high surface area, while \ce{Fe} centers offer strong binding sites via Coulombic \ce{Fe-H2} interactions and favorable spin states \cite{Peng2022}. Theoretical studies reveal that low-spin \ce{Fe^{2+}} centers, as found in ferrocene, can achieve up to a fourfold enhancement in \ce{H2} loading compared to high-spin configurations.\cite{Cha2010}

The iron center can engage in direct coordination interactions, while the cyclopentadienyl rings contribute $\pi$-electron density for additional binding modes. This dual-site accessibility makes ferrocene an attractive candidate for gas storage applications because it can potentially interact with multiple hydrogen molecules simultaneously without compromising structural integrity.

\subsection{Transition metals in COFs: cost considerations}\label{sec:cost}

The incorporation of transition metals into COF structures has been demonstrated to enhance hydrogen storage performance through enhanced adsorbate-framework interactions.\cite{han_recent_2009} However, most successful approaches have relied on expensive precious metals, such as platinum and palladium, limiting practical implementation.\cite{mendoza-cortes_covalent_2012} Iron-based systems offer a compelling alternative that combines favorable electronic properties with abundant availability and low cost. Iron costs approximately \$0.10 per kg compared to platinum at \$40,000+ per kg, representing a 400,000-fold cost advantage at the time of writing.

Previous studies have shown that iron-containing binding sites can achieve hydrogen binding energies in the optimal range of 15--25 kJ$\cdot$mol$^{-1}$, sufficient for room-temperature storage while allowing facile desorption.\cite{Cha2010} The key advantage of ferrocene over other iron-containing species lies in its exceptional stability and well-defined coordination environment, which should maintain consistent binding properties throughout storage-desorption cycles.

\subsection{Design of MSUCOF-4-FeCp}

\begin{figure*}[!htb]
    \centering
    \includegraphics[width=\textwidth]{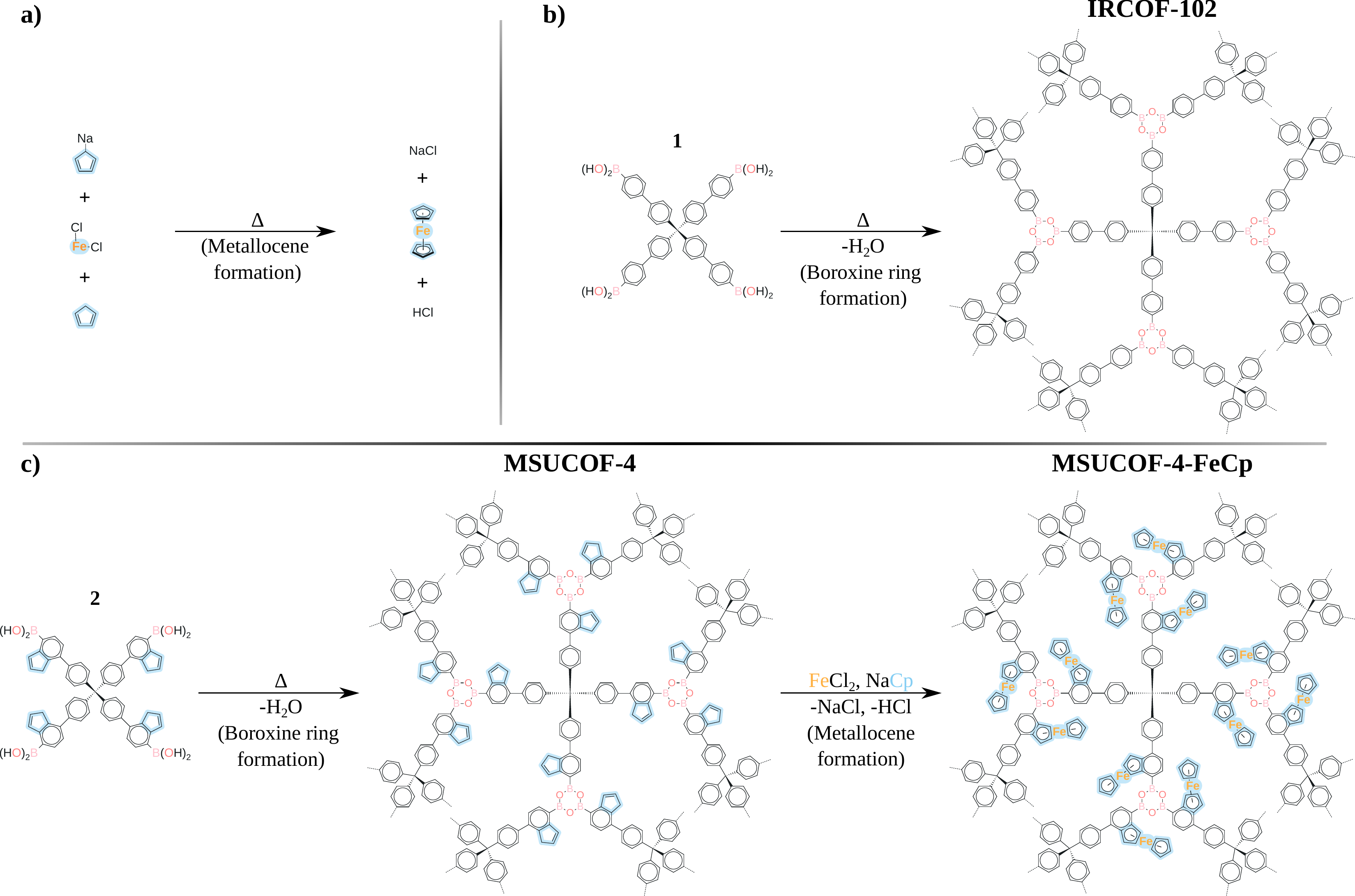}
    \caption{Schematic illustration of framework synthesis and post-synthetic metallation. (a, top left) Representative metallocene formation reaction, \ce{FeCl + Cp + NaCp -> NaCl + FeCp + HCl}. (b, top right) Boroxine condensation of tetrakis(4$'$-borono-[1,1$'$-biphenyl]-4-yl)methane (Linker~1) to form IRCOF-102. (c, bottom) Boroxine ring formation of tetrakis(4-(4-boronoinden-7-yl)phenyl)methane (Linker~2) to yield MSUCOF-4, a modified variant of IRCOF-102 in which biphenyl units are replaced by indene moieties; the cyclopentadienyl (Cp) portion of the indene linker is highlighted in light blue. Subsequent post-synthetic metallocene formation affords the metalated framework MSUCOF-4-FeCp. Atom colours: Fe (orange), B (pink), O (red), and C (black).}
    \label{fig:Formation}
\end{figure*}

In the search for a room temperature storage material, researchers often report gravimetric uptake (weight \%) up to 100 bar, as this metric prioritizes maximum energy density with minimal added weight; a crucial consideration for transportation applications. The evolution of this field is exemplified in Fig. \ref{fig:LitReview}. Early efforts produced materials with exceptional volumetric capacity (Fig. \ref{fig:LitReview} a). The subsequent shift toward prioritizing gravimetric performance has driven exponential improvements among leading MOFs and COFs (Fig. \ref{fig:LitReview} b), yet this optimization often comes at the expense of volumetric performance. This inherent trade-off is illustrated in Fig. \ref{fig:LitReview} c: the most successful materials to date approach either the DOE volumetric or gravimetric target, but not simultaneously both. For example, both taps-COF-1\cite{Xiao-Dong_2019} and tapa-COF-1\cite{Xiao-Dong_2016} successfully exceed the 6.5 wt\% (ultimate gravimetric target) yet achieve less than 20\% of the DOE ultimate volumetric target of 50 g/L. Strong binding sites that enhance volumetric capacity typically require heavy metal centers or dense framework designs that penalize gravimetric metrics. Our ferrocene functionalization strategy circumvents this limitation by introducing highly effective binding sites with minimal mass contribution, enabling MSUCOF-4-FeCp to achieve both volumetric and gravimetric targets; a milestone that to the best of our knowledge has not been previously reported for any metal- or covalent-organic framework.

IRCOF-102 was selected as the baseline framework for ferrocene functionalization due to its robust three-dimensional boroxine-linked topology. As an isoreticular extension\cite{Nguyen2021} of COF-102,\cite{El-Kaderi2007} IRCOF-102 maintains the same tetrahedral node geometry and boroxine chemistry while incorporating extended biphenyl linkers in place of single phenyl units. The framework, constructed from tetrakis(4'-borono-[1,1'-biphenyl]-4-yl)methane (Linker~1), provides larger pores and higher surface area compared to the parent structure while preserving its synthetic accessibility and structural stability that characterize the COF-102 topology. 

Our design strategy employs tetrakis(4-(4-boronoinden-7-yl)phenyl)methane (Linker~2), where the fused cyclopentadienyl ring provides the coordination environment for the formation of ferrocene while the boronic acid at the 4-position enables the condensation of boroxine. This 4,7-substitution maintains a para relationship preserving the linear geometry critical for framework assembly. The framework is designated MSUCOF-4-FeCp instead of MSUCOF-4-\ce{FeCp_2} as a Cp ring is integrated within Linker~2, with only the second Cp ring and the Fe center added post-synthetically.

Ferrocene units are located within tritopic pore regions, where three framework struts converge around each boroxine linkage (Fig.~\ref{fig:Formation}). These tritopic arrangements create cooperative binding environments that enhance binding energies without reaching chemisorption,\cite{Djokic2024} with ferrocene moieties providing multiple binding sites (Fe centers and Cp rings) for complementary hydrogen interactions.

The preferred synthetic route involves the condensation of Linker~2 to form MSUCOF-4, followed by post-synthetic metallation with \ce{FeCl2} and \ce{NaCp} to produce MSUCOF-4-FeCp. This approach enables comprehensive characterization before metallation and avoids the potential coordination of iron with boroxine-forming groups during synthesis. The transformation proceeds through the deprotonation of the cyclopentadiene ring already in MSUCOF-4 to generate \ce{Cp-}, which coordinates with \ce{Fe^{2+}} and an additional \ce{Cp-} (from NaCp) to form the ferrocene unit without sterically hindering pore accessibility. The post-synthetic strategy offers superior control over both framework formation and metal incorporation, consistent with successful metallation protocols in related COF systems.\cite{Dong2022}

\section{Computational methods}

\subsection{Quantum mechanical calculations}

Density functional theory (DFT) calculations were performed to determine the electronic structure, binding properties, and optimal geometries of ferrocene-functionalized fragments and the complete MSUCOF-4-FeCp framework. Workflow automation for CRYSTAL23 calculations was managed using our MACE (Mendoza Automated CRYSTAL Engine) python package,\cite{MACE_zenodo} an open-source framework for automating CRYSTAL quantum chemistry workflows that include input generation, queue management, and error recovery. MACE is freely available to enable other researchers to easily adopt and extend these computational workflows.

\subsubsection{Fragment calculations for force field development}\label{sec:fragDFT}

Fragment calculations were performed using CRYSTAL23\cite{CRYSTAL23,CRYSTAL23_2} with the M06 functional\cite{M06} and the pob-TZVP-rev2 basis set\cite{TZVP-rev2} employing spin-polarized treatment with an extra-large integration grid.\cite{Laun2022} The M06 functional was selected for its balanced performance in reproducing thermochemistry and non-covalent interactions.\cite{Mardirossian} Geometry optimizations employed convergence criteria of $2.72 \times 10^{-11}$ eV for SCF energy, $1.54 \times 10^{-3}$ eV \AA$^{-1}$ for RMS forces, and $6.35 \times 10^{-5}$ \AA\ for maximum atomic displacements, with $\Gamma$-point only Monkhorst-Pack sampling of the 500 \AA\ supercell.

\subsubsection{Periodic framework calculations} \label{sect:PeriodicDFT}

Full periodic calculations used CRYSTAL23 with HSE06-D3\cite{HSE06_2003,HSE06_2006,d3} providing accurate treatment of electronic properties such as band gap, band structure, and density of states.\cite{borlido_exchange-correlation_2020} Comprehensive geometric optimizations for atomic positions and lattice parameters used identical convergence criteria as \ref{sec:fragDFT} with Monkhorst-Pack $\Gamma$-centered grids. Notably, ferrocene units adopt eclipsed Cp conformations within the framework, contrary to the staggered preference of isolated molecules, because of spatial constraints and electronic interactions with neighboring struts. This eclipsed geometry was maintained in all force field parameterizations fitting. The optimized fractional coordinates and lattice parameters for IRCOF-102, MSUCOF-4, and MSUCOF-4-FeCp are reported in the SI Section \ref{sec:cif}.

\subsection{Force field development and validation}

Quantum mechanics-derived force fields were developed following our previous methodology\cite{Djokic2024} employing a Morse potential (Equation~\ref{eq:Morse}) for \ce{H2}-ferrocene interactions:

\begin{equation} \label{eq:Morse}
U_{ij}^{\text{Morse}} (r_{ij}) = D_0[(1-\text{e}^{-\alpha(r_{ij}-r_{0})})^2-1] 
\end{equation}

\begin{figure}[h]
\centering
\includegraphics[width=\columnwidth]{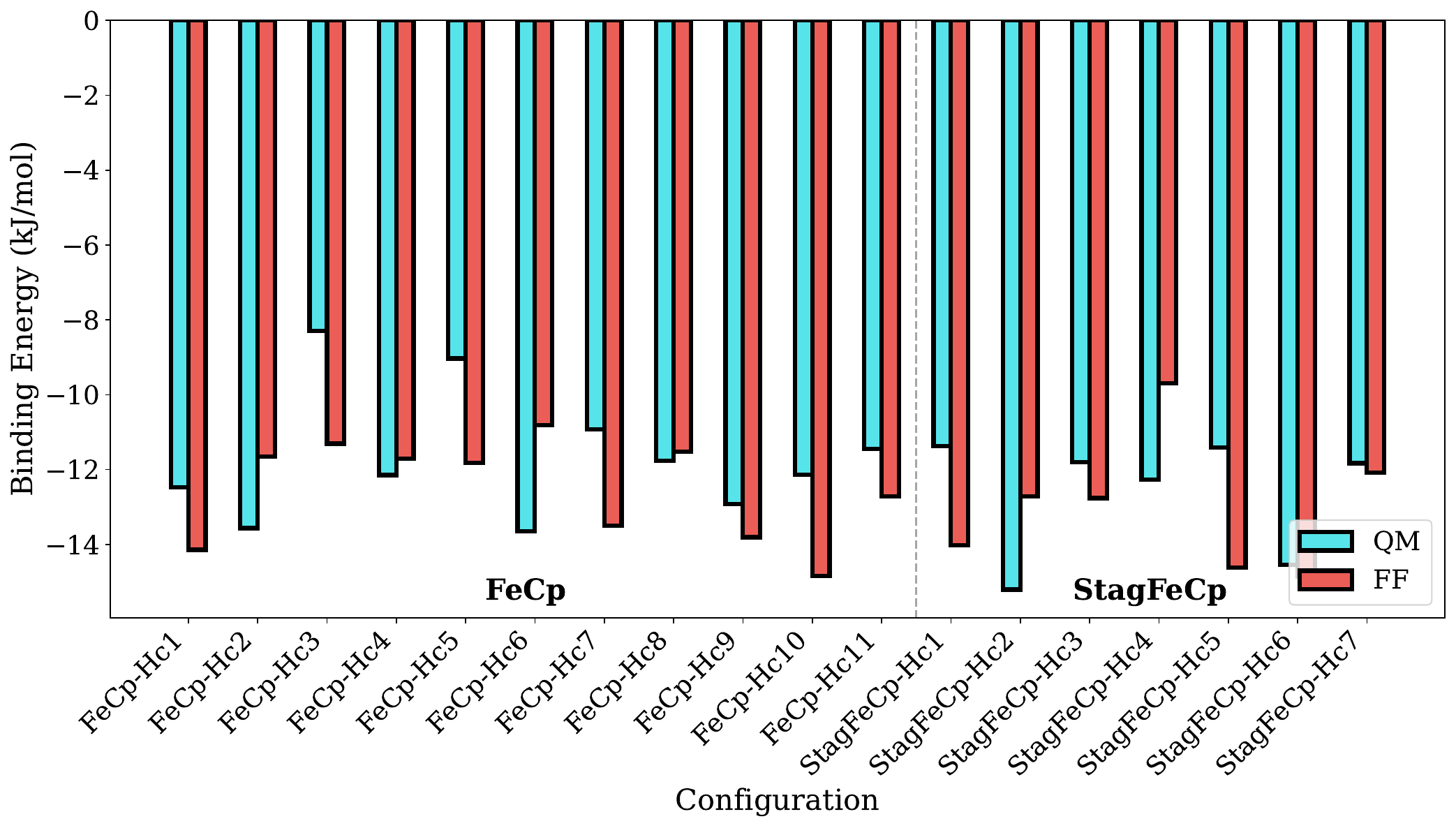}
\caption{Validation of the QM-fitted force field for ferrocene--\ce{H2} interactions. Hc denotes hydrogen configuration number. FeCp labels correspond to eclipsed ferrocene (as observed in MSUCOF-4-FeCp), while stagFeCp denotes staggered conformers.}
\label{fig:ff_validation}
\end{figure}

The parameters ($D_0$, $\alpha$, $r_0$) were fitted to DFT binding energy curves using GULP,\cite{GULP} sampling multiple Hydrogen configurations (Hc) around eclipsed and staggered variants of ferrocene. Representative training configurations are shown in Fig. \ref{fig:configurations}. A least-squares fitting procedure achieved mean absolute errors below 0.5 kJ$\cdot$mol$^{-1}$ for test configurations (Fig. \ref{fig:ff_validation}).

\begin{table}[h]
\small
\centering
\begin{tabular}{lccc}
\hline
\textbf{Interaction} & \textbf{$D_0$ (kcal$\cdot$mol$^{-1}$)} & \textbf{$\alpha$ (\AA$^{-1}$)} & \textbf{$r_0$ (\AA)} \\
\hline
\ce{H2}-\ce{C} & 0.101 & 1.924 & 3.120 \\
\ce{H2}-\ce{H} & 0.001 & 1.849 & 3.247 \\
\ce{H2}-\ce{Fe} & 1.987 & 0.466 & 2.476 \\
\hline
\end{tabular}
\caption{Morse potential parameters for \ce{H2}-ferrocene interactions derived from DFT calculations.}
\label{tab:morse_params}
\end{table}

The optimized parameters shown in Table \ref{tab:morse_params} reveal distinctive binding characteristics for the ferrocene iron center. The \ce{H2}-\ce{Fe} interaction exhibits a notably deep potential well ($D_0$ = 1.987 kcal$\cdot$mol$^{-1}$) combined with a small width parameter ($\alpha$ = 0.466 \AA$^{-1}$), indicating diffuse and extended interaction. This combination produces a broad, attractive potential that facilitates hydrogen capture while maintaining reversibility. In contrast, the chelation complex \ce{FeCl2} of our prior study exhibited drastically different electronic environment ($D_0$ = 1.092 kcal$\cdot$mol$^{-1}$, $\alpha$ = 1.180 \AA$^{-1}$, $r_0$ = 3.155 \AA), with weaker, shorter-range binding despite greater metal accessibility.\cite{Djokic2024} Notably the cyclopentadienyl rings geometrically constrain the Fe center of ferrocene, yet donate substantial $\pi$-electron density which, in addition to its low spin electronic configuration, creates an electron-rich environment that strengthens Coulombic interaction with the \ce{H2} quadrupole moment.

\subsection{Grand Canonical Monte Carlo simulations}

Hydrogen storage isotherms were calculated using Grand Canonical Monte Carlo (GCMC) simulations implemented in Materials Studio\cite{biovia_materials_2022_updated} at 298 K and 1--700 bar. The Metropolis algorithm used translation, rotation, insertion, deletion, and regrowth moves in a 2:1:1:1:0.1 ratio. Equilibrium was achieved after 1,000,000 steps, followed by 3,000,000 production steps to average uptake and isosteric heat of the ensemble ($Q_{\text{st}}$).

\subsection{Equation of state}

To convert between fugacity (used in GCMC simulations) and pressure (for experimental comparison), the van der Waals equation of state was employed:\cite{winn_fugacity_1988}

\begin{equation} \label{eq:vdw}
\ln \frac{f}{P} = \left(b - \frac{a}{RT}\right)\frac{P}{RT}
\end{equation}

\noindent where $f$ is the fugacity, $P$ is the pressure, $T$ is the temperature (298 K), $R$ is the gas constant, and $a$ = 0.2476 L$^2$ bar mol$^{-2}$ and $b$ = 0.02661 L mol$^{-1}$ are empirical parameters for hydrogen. In our previous work, we showed that this formulation showed only 0.735\% difference from Peng-Robinson at 700 bar.\cite{Djokic2024}

\begin{figure*}[!htb]
\centering
\includegraphics[width=\textwidth]{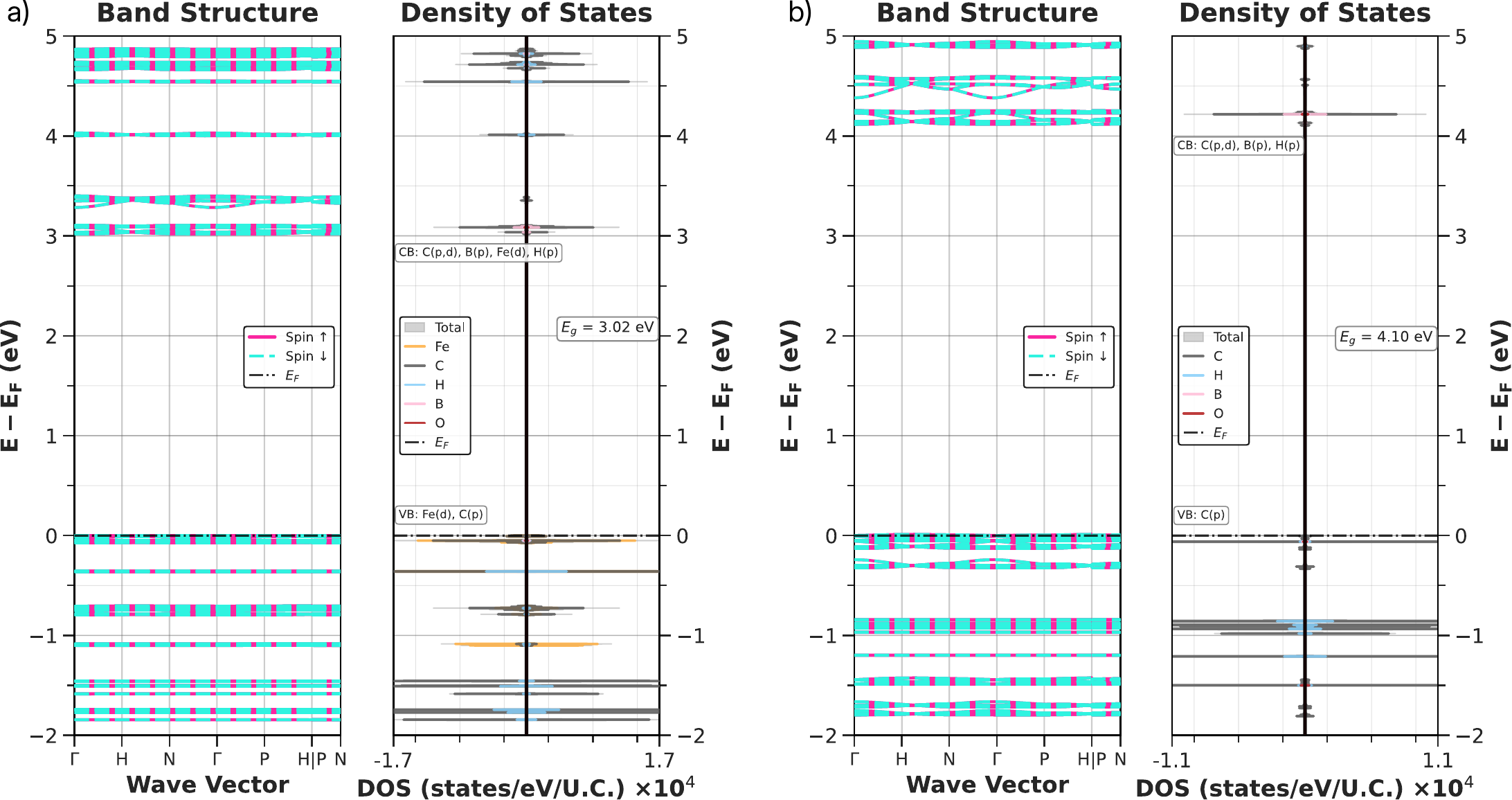}
\caption{Electronic structure of (a) MSUCOF-4-FeCp and (b) IRCOF-102, showing band structure (left) and density of states (right). Both materials exhibit flat bands typical of porous materials. Functionalization in MSUCOF-4-FeCp narrows the band gap, enhancing the potential for electronic or catalytic applications.} 
\label{fig:BandDoss}
\end{figure*}

\section{Results and discussion}

\subsection{Structural properties and design validation}

Incorporation of ferrocene into the IRCOF-102 framework results in MSUCOF-4-FeCp, which has well-defined structural properties that support high hydrogen storage performance. As shown in Table~\ref{tab:cof_comparison}, MSUCOF-4-FeCp maintains a high surface area of 4780 m$^2$ g$^{-1}$ while achieving a balanced pore volume of 2.55 cm$^3$ g$^{-1}$ and a void fraction of 0.84. The framework density of 0.12 g cm$^{-3}$ is significantly lower than the precious metal-functionalized variants, contributing to enhanced gravimetric performance. For comparison, structural and uptake results for MSUCOF-(1-3) from our previous work\cite{Djokic2024} are included. Surface areas and pore volumes were determined using a hydrogen-sized probe rolling across the van der Waals surface of each framework (Fig. \ref{fig:vdw}; see SI Section \ref{sec:SA} for details)

\begin{table}[h]
\small
\caption{Comparison of geometric properties of selected MSUCOFs}
\label{tab:cof_comparison}
\begin{tabular*}{\columnwidth}{@{\extracolsep{\fill}}lcccc}
\hline
COF name & \begin{tabular}{@{}c@{}}Surface area\\(m$^2$ g$^{-1}$)\end{tabular} & \begin{tabular}{@{}c@{}}Pore volume\\(cm$^3$ g$^{-1}$)\end{tabular} & \begin{tabular}{@{}c@{}}Void\\fraction\end{tabular} & \begin{tabular}{@{}c@{}}Density\\(g cm$^{-3}$)\end{tabular} \\
\hline
MSUCOF-1 & 5060 & 5.06 & 0.91 & 0.07 \\
MSUCOF-1-PtCl & 2890 & 2.09 & 0.88 & 0.15 \\
MSUCOF-3-CoCl & 3770 & 0.94 & 0.73  & 0.28 \\
MSUCOF-3-PtCl & 2190 & 0.55 & 0.72 & 0.47 \\
IRCOF-102 & 5680 & 5.11 & 0.90 & 0.06 \\
MSUCOF-4 & 4840 & 4.14 & 0.88 & 0.08 \\
MSUCOF-4-FeCp & 4780 & 2.55 & 0.84 & 0.12 \\
\hline
\end{tabular*}
\end{table}

Pristine COFs often suffer from poor binding energies at room-temperature. Strategies to address this limitation include functionalization, such as incorporating ferrocene onto the ligands. Not only does this strategy inject strong binding sites throughout the pore, it does so without an excessive binding strength that can compromise the deliverable capacity (see Section \ref{sect:Binding}).

It is important to note that MSUCOF-4 incorporates one cyclopentadienyl ring as an extended variant of the linkers present in IRCOF-102, yet it lacks the complete Fe($\eta^5$-C$_5$H$_5$) moiety found in MSUCOF-4-FeCp. This partial functionalization accounts for the intermediate geometric properties observed: MSUCOF-4 shows reduced surface area (4840 vs 5680 m$^2$ g$^{-1}$), pore volume (4.14 vs 5.11 cm$^3$ g$^{-1}$), and void fraction (0.88 vs 0.90) compared to pristine IRCOF-102. However, as shown in subsequent sections, MSUCOF-4 still exhibits hydrogen storage performance comparable to IRCOF-102 despite these geometric differences. This indicates that the exceptional performance of MSUCOF-4-FeCp is primarily due to the ferrocene complex-hydrogen interactions rather than purely geometric factors.

\subsection{Electronic structure calculations}

The electronic properties of the pristine IRCOF-102 and functionalized MSUCOF-4-FeCp were studied and visualized by calculating the band structure and the density of states (Fig. \ref{fig:BandDoss}). For these calculations, we follow the methods described in Section \ref{sect:PeriodicDFT}. These calculations were performed using spin unrestricted DFT, allowing $\alpha$ and $\beta$ orbitals to vary to capture any potential magnetic ordering. These MSUCOF-4 and IRCOF-102 families of materials follow the \textit{I}-$\bar{4}3$\textit{d} space group (No. 220); accordingly the reciprocal space was sampled in the high symmetry path \textbf{$\Gamma$-H-N-$\Gamma$-P-H$|$P-N} as determined by SeeKPath to capture the relevant band dispersion.\cite{SeeKPath} The atom-projected density of states will reveal the atomic contributions at frontier bands. 

These results show expected trends for porous materials. In both materials, many of the bands exhibit minimal band dispersion (flat bands with no spin asymmetry), indicative of localized molecular orbitals throughout the organic linkages. Additionally, IRCOF-102 displays a band gap large enough to classify it as an insulator ($E_g = 4.10$ eV), which is consistent with its composition of light main-group elements (C, B, O) engaged in saturated covalent bonding. MSUCOF-4-FeCp acts as a functionalized variant with Fe($\eta^5$-C$_5$H$_5$)$_2$ additions in tritopic regions within the pore. By adding this moiety, MSUCOF-4-FeCp experiences a more narrow band gap of $E_g = 3.02$ eV (notably ~410 nm in the visible light range), enhancing its potential for electronic and catalytic applications. 

\begin{figure*}[!htb]
\centering
\includegraphics[width=0.8\textwidth]{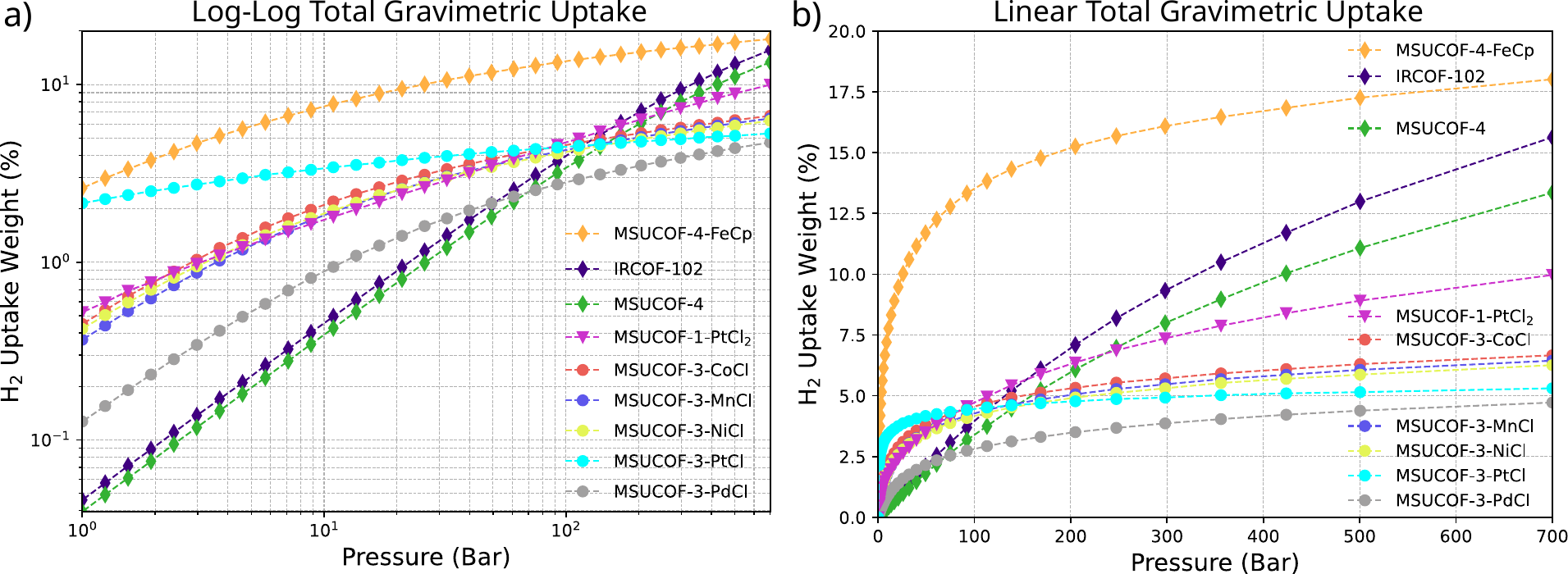}
\vskip 1 mm 
\includegraphics[width=0.8\textwidth]{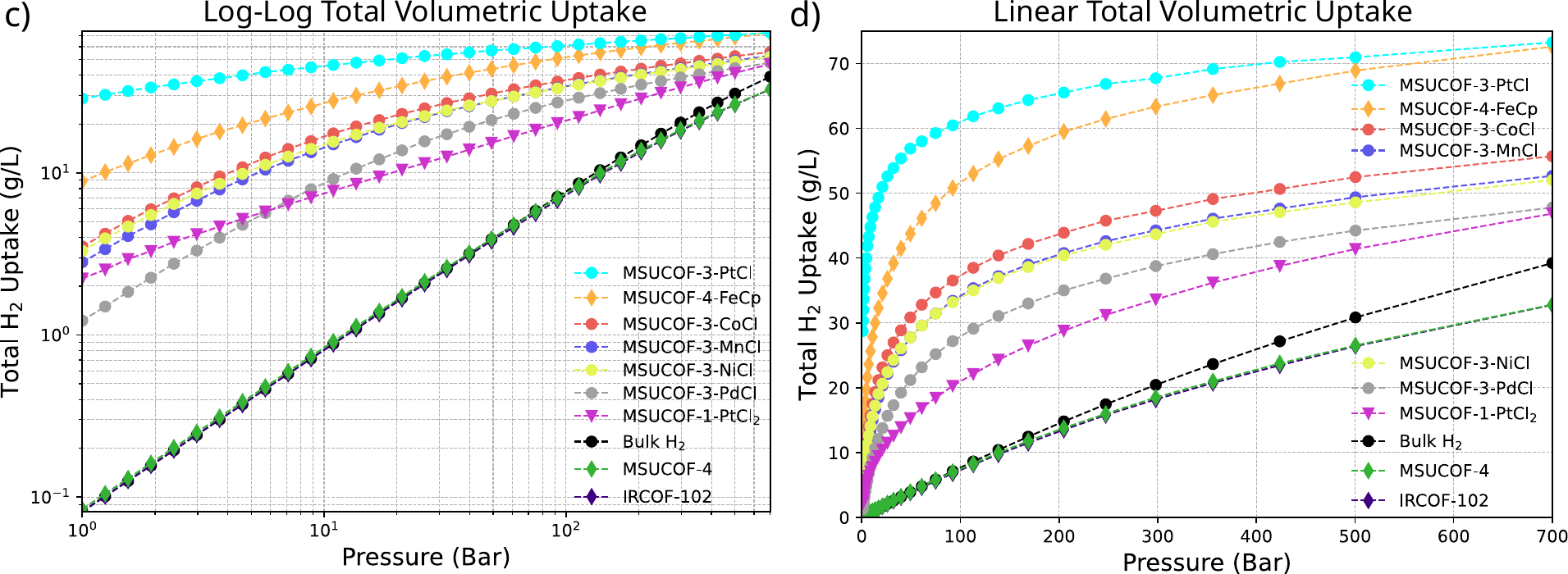}
\caption{Comparison of total gravimetric (a,b) and volumetric (c,d) uptake at 298 K for the MSUCOF-4 family against top-performing MSUCOF variants. Log-log plots (a,c) provide detail in the low-pressure region, while linear plots (b,d) show high-pressure behavior. MSUCOF-4-FeCp demonstrates superior performance across all pressure ranges.} % Remember to update subtitle to log-log and maybe linea/stanard
\label{fig:performance_isotherms}
\end{figure*}

Analysis of the HOMO and LUMO character of MSUCOF-4-FeCp reveals that the valence band edge is predominantly Fe-centered, with contributions from Fe(d) and C(p) orbitals of the cyclopentadienyl rings. Meanwhile, the conduction band edge is localized on an organic framework, primarily C(p,d), with minor contributions from B(p), Fe(d), and H(p). This spatial separation of frontier orbitals is indicative of metal-to-ligand charge transfer, suggesting that photoexcitation would generate oxidized Fe centers and reduced organic sites. These charge-separated states are favorable for both photocatalysis and electrocatalysis, as is common with ferrocene-based systems. The flat bands reflect the localized nature of these redox-active sites, which is typical of molecular catalysts embedded in porous frameworks and does not preclude efficient catalysis where electron transfer could occur at discrete metal centers rather than in a band-like transport.

\subsection{Hydrogen storage performance}

The hydrogen storage performance of MSUCOF-4-FeCp demonstrates exceptional promise in meeting DOE targets. Fig. \ref{fig:performance_isotherms} presents both gravimetric and volumetric uptake isotherms at 298 K, showing that MSUCOF-4-FeCp achieves 18.0 wt\% hydrogen uptake and 72.6 g \ce{H2} L$^{-1}$ at 700 bar. This performance represents a significant improvement over the parent IRCOF-102, and the gravimetric uptake is approximately 3.5 times higher than that of the denser MSUCOF-3-PtCl variant. We chose to measure up to these pressures as existing commercial hydrogen fuel cell vehicles are already available to the market offering pressure tanks up to 700 bar.

Working capacity (WC)—defined as the amount of hydrogen that can be practically delivered between high storage pressure (700 bar) and rough delivery pressure (5 bar)—is a critical heuristic to approximate realistic delivery that can be obtained in a reasonable time frame. Unlike MSUCOF-3-PtCl, which detrimentally locks hydrogen even at low pressures, MSUCOF-4-FeCp maintains a balanced adsorption profile that ensures higher usable capacity under working conditions. 

The log-log total uptake plots (Fig. \ref{fig:performance_isotherms} a,c) best capture low pressure adsorption behavior, where the linear response reflects the Henry regime in which uptake scales proportionally with pressure. Sub-linear behavior develops as saturation effects emerge, ultimately reaching a plateau that is characteristic of Langmuir-like saturation. In contrast, the linear plots (Fig. \ref{fig:performance_isotherms} b,d) better resolve high pressure behavior. These complementary representations reveal distinct material characteristics: strong adsorbents such as MSUCOF-3-PtCl dominate at low pressure, but also saturate earlier.

In terms of volumetric uptake (Fig. \ref{fig:performance_isotherms} b), MSUCOF-4-FeCp retains less \ce{H2} at low pressures than MSUCOF-3-\ce{PtCl2}, but still more than the first-row transition metal (TM) chelation complexes of MSUCOF-3-TM\ce{Cl2}. However, the linear volumetric uptake plot (Fig. \ref{fig:performance_isotherms} d) reveals that MSUCOF-4-FeCp approaches the energy density of MSUCOF-3-\ce{PtCl2} at high pressures. This balance is critical for volumetric working capacities (WC$_v$): MSUCOF-4-FeCp achieves 52.2 g \ce{H2} L$^{-1}$, surpassing all previous variants of MSUCOF-1–3 (32.6–44.3 g \ce{H2} L$^{-1}$) by accumulating substantial \ce{H2} at high pressures without detrimentally locking it in low pressures. Despite its exceptional total volumetric uptake, MSUCOF-3-\ce{PtCl2} hinders its own WC$_v$ to 32.6 g \ce{H2} L$^{-1}$ precisely for this reason.

\begin{figure*}[!htbp]
\centering
\includegraphics[width=0.8\textwidth]{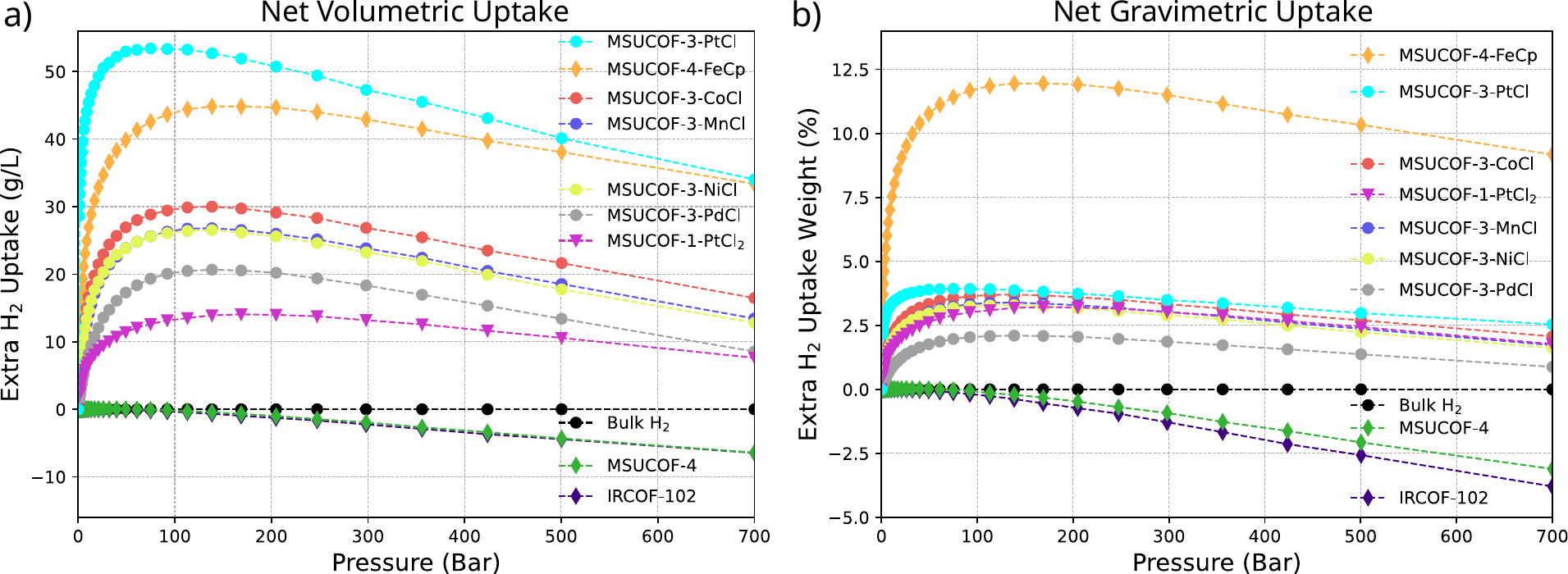}
\caption{High-pressure \ce{H2} net volumetric and gravimetric isotherms showing deliverable capacity advantages of MSUCOF-4-FeCp compared to other variants at 298 K.}
\label{fig:net_uptake}
\end{figure*}

The spacious unit cell inherited from IRCOF-102 (roughly $4\times$ the volume of MSUCOF-3) reduces the density of the framework, allowing MSUCOF-4-FeCp to achieve its high volumetric performance without sacrificing its gravimetric uptake. Gravimetric uptake increases from 2.6 wt\% at 1 bar to 18.0 wt\% at 700 bar, yielding a gravimetric working capacity (WC$_g$) of 12.2 wt\%. The low-density MSUCOF-2 family achieves 11.4-18.9 wt\% WC$_g$ but suffers from poor volumetric performance (33.2-38.7 g \ce{H2} L$^{-1}$), while the pristine IRCOF-102 (15.4 wt\%, 32.4 g \ce{H2} L$^{-1}$) closely matches the pristine MSUCOF-1 (15.2 wt\%, 32.4 g \ce{H2} L$^{-1}$). Thus, MSUCOF-4-FeCp is one of the only materials capable of simultaneously satisfying both DOE ultimate targets: 6.5 wt\% gravimetric capacity and 50 g \ce{H2} L$^{-1}$ volumetric capacity.

The net uptake metric provides another crucial performance indicator by comparing the porous material with compressed bulk hydrogen under identical conditions. Positive net uptake values indicate that the incorporation of the porous material provides a storage advantage over simple compression, while negative values suggest that the added weight and volume of the material actually reduce overall storage efficiency. Excess uptake isotherms, which apply a less stringent correction using only the accessible pore volume, are provided in Fig. \ref{fig:excess_isotherm}.

The net uptake plot (Fig. \ref{fig:net_uptake}) reveals peaks at optimal pressures ($P_{\text{opt}}$) that reflect the interplay between binding strength and saturation behavior. MSUCOF-4-FeCp achieves $P_{\text{opt}}$ = 145 bar; substantially higher than MSCUOF-3-PtCl ($P_{\text{opt}}$ = 65 bar), which saturates prematurely due to its stronger binding and smaller pores, yet comparable to the dense first-row transition metal variants of MSUCOF-3 ($P_{\text{opt}}$ $\approx$ 130-139 bar). Crucially, MSUCOF-4-FeCp maintains positive net uptake across a broad pressure range, validating that the enhanced ferrocene binding interactions more than compensate for the framework's mass and volume even at 700 bar. These metrics are summarized in Table
\ref{tab:working_capacity}. Based on this usable working capacity, several variants of MSUCOF achieve either the 2025 DOE gravimetric or volumetric target, but MSUCOF-4-FeCp is still the only material that simultaneously satisfies both ultimate targets (6.5 wt\% and 50 g \ce{H2} L$^{-1}$).

\begin{table}[htbp]
\centering
\caption{Working capacity performance metrics for select MSUCOF variants at 298 K. $P_{\text{opt}}$ denotes the optimal pressure at which net uptake is maximized, WC$_g$ and WC$_v$ represent gravimetric and volumetric working capacities (700 bar to 5 bar), respectively.}
\label{tab:working_capacity}
\begin{tabular}{lccc}
\hline
\textbf{Material} & $\boldsymbol{P_{\textbf{opt}}}$ & \textbf{WC}$_{\boldsymbol{g}}$ & \textbf{WC}$_{\boldsymbol{v}}$ \\
 & (bar) & (wt\%) & (g \ce{H2} L$^{-1}$) \\
\hline
MSUCOF-1-PtCl$_2$ & 169 & 8.71 & 41.4 \\
MSUCOF-3-CoCl & 139 & 5.22 & 44.3 \\
MSUCOF-3-MnCl & 139 & 5.20 & 43.1 \\
MSUCOF-3-NiCl & 130 & 4.95 & 41.7 \\
MSUCOF-3-PdCl & 139 & 4.20 & 42.7 \\
MSUCOF-3-PtCl & 75 & 2.29 & 32.6 \\
IRCOF-102 & 3 & 15.4 & 32.4 \\
MSUCOF-4 & 26 & 13.2 & 32.8 \\
\textbf{MSUCOF-4-FeCp} & \textbf{145} & \textbf{12.2} & \textbf{52.2} \\
\hline
\multicolumn{4}{l}{\textit{DOE Targets}} \\
2025 & -- & 5.5 & 40.0 \\
Ultimate & -- & 6.5 & 50.0 \\
\hline
\end{tabular}
\end{table}
\vspace{-5 mm}

\subsection{Binding energy analysis and mechanisms\label{sect:Binding}} 

\begin{figure}[!htb]
\centering
\includegraphics[width=0.8\columnwidth]{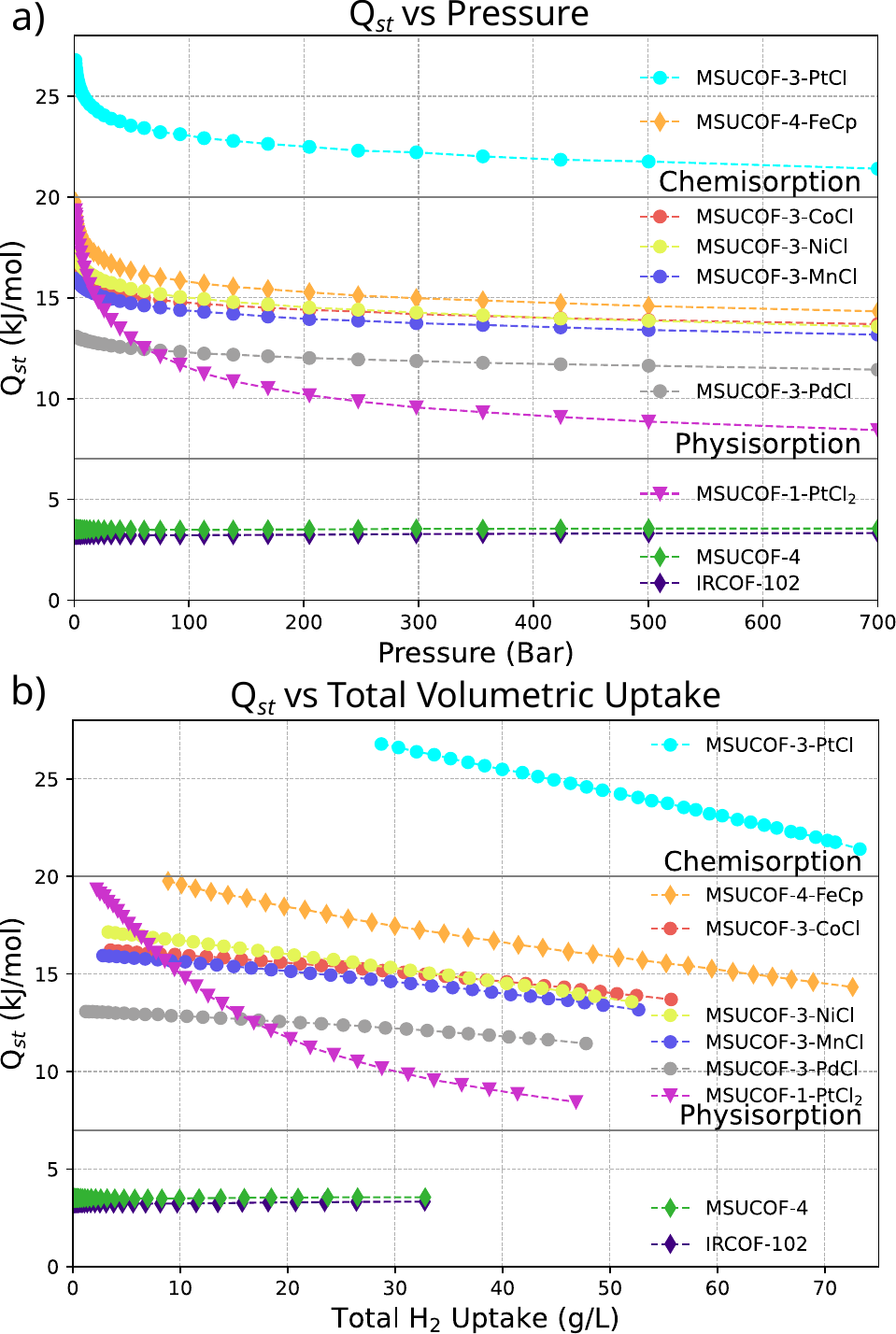}
\caption{High-pressure \ce{H2} isosteric heat of adsorption ($Q_{\text{st}}$) at 298 K versus (a) Pressure and (b) total \ce{H2} uptake, showing optimal binding energy characteristics for MSUCOF-4-FeCp, which lies entirely between chemi- and physisorption for \ce{H2}.}
\label{fig:qst_analysis}
\end{figure}

Analysis of isosteric heat of adsorption ($Q_{\text{st}}$) demonstrates that MSUCOF-4-FeCp achieves optimal binding energies within the intermediate physisorption regime (Fig. \ref{fig:qst_analysis}). Interestingly, the binding energy profile is similar to that of MSUCOF-3-CoCl, a material that we previously recognized for its superior working capacity. Moderate binding strengths (7--20 kJ$\cdot$mol$^{-1}$) allow efficient hydrogen adsorption at high pressures without locking hydrogen into the structure.

 % maybe add background color for chemi and physi sorption. Add transparency for chemi and physi text

From the GCMC simulations, we can derive a probability distribution $P(E)$ that illustrates the proportions of adsorbed \ce{H2} molecules experiencing interaction energy $E$ at thermodynamic equilibrium. This distribution satisfies the normalization condition:

\begin{equation}
\int_{-\infty}^{\infty} P(E) \, dE = 1
\label{eq:normalization}
\end{equation}

In Equation \ref{eq:normalization}, the negative integration limits are associated with energies that correspond to attractive interactions, while any interaction greater than zero is repulsive. The isosteric heat of adsorption is directly tied to the ensemble-averaged interaction energy through:

\begin{equation}
Q_{\text{st}} = -\langle E \rangle = -\int_{-\infty}^{\infty} E \cdot P(E) \, dE
\label{eq:qst_average}
\end{equation}

This relationship (Equation \ref{eq:qst_average}) shows that $Q_{\text{st}}$ represents the mean of the energy distribution. Note the negative sign that converts the interaction energy to a positive heat of adsorption, following the standard reporting convention. Then it follows that the variance of the distribution can be calculated with the following.

\begin{equation}
\sigma^2 = \int_{-\infty}^{\infty} (E - \langle E \rangle)^2 \cdot P(E) \, dE
\label{eq:variance}
\end{equation}

Equation \ref{eq:variance} quantifies the diversity of binding sites within the framework. A larger variance indicates a broader range of binding environments, which can be advantageous for maintaining adsorption across various pressure levels.

\begin{figure}[!htbp]
\centering
\includegraphics[width=0.97\columnwidth]{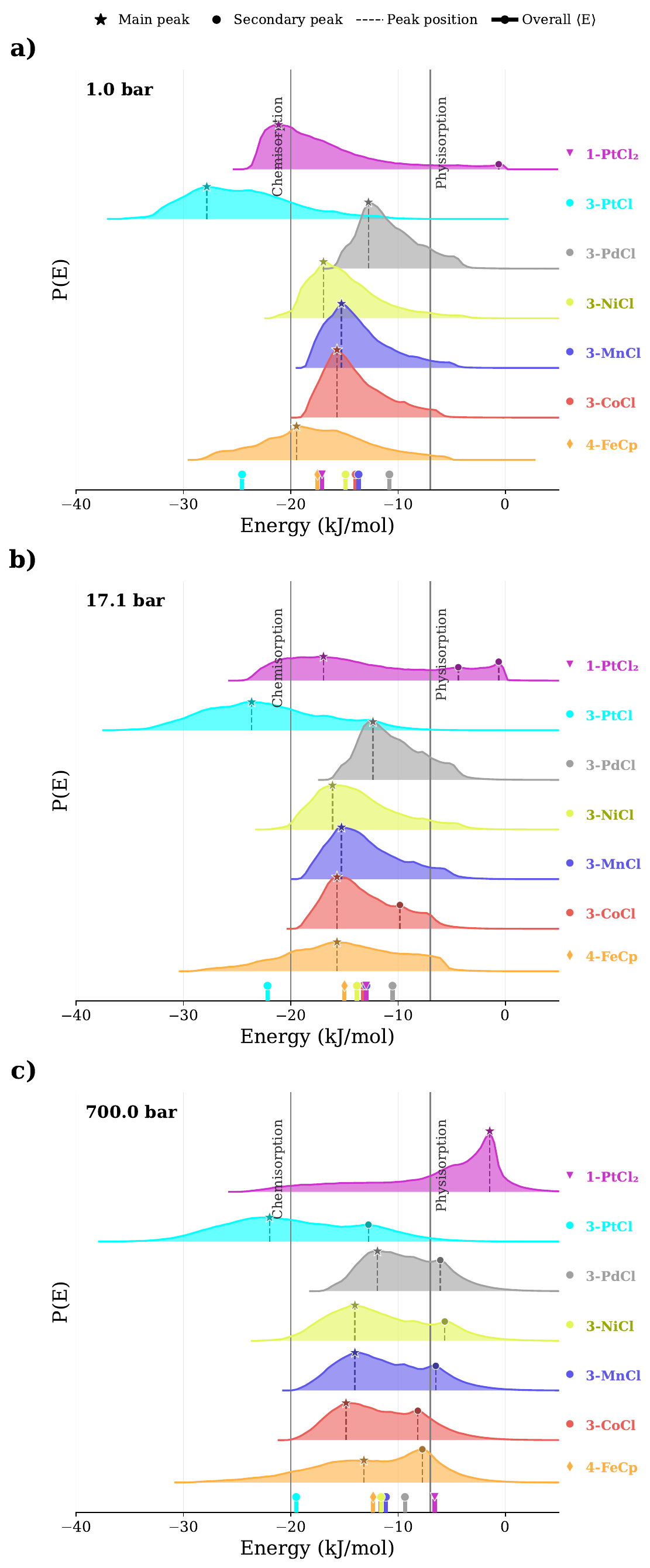}
\caption{Ridge plot of \ce{H2} interaction energy distributions $P(E)$ at (a) 1, (b) 17.1, and (c) 700 bar; curves are vertically offset for clarity. Solid stars and open circles mark primary and secondary $Q_{\text{st}}$ peaks, respectively. Colored ticks indicate ensemble-averaged energies $\langle E \rangle$; vertical 
dashed lines delineate chemisorption ($>$20 kJ$\cdot$mol$^{-1}$) and physisorption ($<$7 kJ$\cdot$mol$^{-1}$) regimes.}
\label{fig:energy_dist}
\end{figure}

The analysis of the energy distribution shed light on the binding mechanisms within select MSUCOFs (Fig. \ref{fig:energy_dist}). At 1 bar, the distribution shows a primary broad peak with a long tail towards 0, indicating that the majority of initial interactions occur at favorable primary binding sites, while few stragglers permeate the pore and interact at weaker sites. At higher pressures, there is a clear right shift of both the overall average binding strength ($Q_{\text{st}}$) and the average of each individual site ($\mu_i$) as they begin to saturate. When secondary peaks appear as shoulders on the primary distribution, accurate quantification requires a method to de-convolute the isolated individual contributions. Each peak $i$ is fitted with an asymmetric (skewed) Gaussian function:

\begin{equation}
P_i(E) = 
\begin{cases}
A_i \exp\left(-\frac{(E - \mu_i)^2}{2\sigma_{L,i}^2}\right) & \text{for } E \leq \mu_i \\[10pt]
A_i \exp\left(-\frac{(E - \mu_i)^2}{2\sigma_{R,i}^2}\right) & \text{for } E > \mu_i
\end{cases}
\label{eq:skewed_gaussian}
\end{equation}

\noindent where $A_i$ is the amplitude, $\mu_i$ is the center (mean energy), and $\sigma_{L,i}$ and $\sigma_{R,i}$ are the widths left and right, respectively. This asymmetric form captures the characteristic skewness of the interaction energy distributions. All peaks are fitted simultaneously using a composite model to avoid double-counting in overlapping regions while preserving an accurate determination of individual binding site energies. In Equation \ref{eq:composite}, $N$ is the number of peaks and all parameters.

\begin{equation}
P(E) = \sum_{i=1}^{N} P_i(E)
\label{eq:composite}
\end{equation}

The average energy for each de-convoluted peak is given by its fitted Gaussian mean $\mu_i$:

\begin{equation}
\langle E \rangle_{\text{peak},i} = \mu_i
\label{eq:peak_avg_energy}
\end{equation}

This approach provides site-specific energies that are more accurate than simple numerical integration over peak boundaries, which can be biased by overlapping contributions from adjacent peaks. The fraction of \ce{H2} molecules that experience energies within the peak $i$ is shown in Equation \ref{eq:site_fraction}. The individual Gaussian component $P_i(E)$ is integrated throughout the energy range and normalized by the total area under $P(E)$ to determine the fraction of molecules that experience that particular binding mode.

\begin{equation}
f_{\text{peak},i} = \frac{\int_{-\infty}^{\infty} P_i(E) \, dE}{\int_{-\infty}^{\infty} P(E) \, dE}
\label{eq:site_fraction}
\end{equation}

 % change color of the legend to black, maybe repeat it at the b and c positions. Add space between 4-FeCp and the overall average, 

At 1 bar, the main peak for MSUCOF-4-FeCp, which is centered around $-18.8$ kJ$\cdot$mol$^{-1}$ ($-4.5$ kcal$\cdot$mol$^{-1}$), highlights the dominant Fe-\ce{H2} interaction. Any unusual features at weaker binding energies likely stem from any of the following: geometries that are sterically constrained; shielded \ce{H2}-\ce{H2} interactions; compounded effects at key overlapped regimes amongst tritopic regions; or interactions with the cyclopentadienyl rings.

When pressure increases to 700 bar, the distribution shows two distinct peaks at $-13.2$ and $-7.7$ kJ$\cdot$mol$^{-1}$ ($-3.2$ and $-1.8$ kcal$\cdot$mol$^{-1}$), which relate to different binding environments within the ferrocene coordination sphere. The predominant broad peak at $-13.2$ kJ$\cdot$mol$^{-1}$ likely corresponds to coordination with the low-spin \ce{Fe^{2+}} center in ferrocene, which theoretical studies have shown provides enhanced \ce{H2} binding compared to high-spin configurations.\cite{Cha2010} Likewise, the sharp peak at $-7.7$ kJ$\cdot$mol$^{-1}$ can be partially attributed to previously mentioned weaker binding regimes that serve as a backup when the primary binding modes saturate.

As pressure evolves, $P(E)$ shows a sequential occupation of binding sites based on their relative affinities. At low pressures, mainly the strongest sites (leftmost region of $P(E)$) are populated, resulting in higher average binding energies. As pressure increases and strong sites saturate, weaker sites begin to dominate, shifting the distribution toward less negative energies and reducing the ensemble-averaged $Q_{\text{st}}$.

From the definition of $Q_{\text{st}} = -\langle E \rangle$ in Equation \ref{eq:qst_average}, the response of the binding energy upon loading follows directly:

\begin{equation}
\frac{\partial Q_{\text{st}}}{\partial n} = -\frac{\partial \langle E \rangle}{\partial n}
\label{eq:qst_loading}
\end{equation}

\noindent where $n$ represents the hydrogen load. For heterogeneous adsorbents possessing a distribution of binding site energies, adsorption should proceed sequentially: the strongest sites populate first, followed by progressively weaker sites as loading increases. Consequently, $\langle E \rangle$ shifts towards less negative values with increasing $n$. In contrast, homogeneous adsorbents with uniform binding sites remain constant regardless of loading.

% \clearpage

\begin{equation}
\frac{\partial Q_{\text{st}}}{\partial n} 
\begin{cases}
< 0 & \text{(heterogeneous)} \\
\approx 0 & \text{(homogeneous)}
\end{cases}
\label{eq:qst_adsorbent_type}
\end{equation}
 
 This negative slope is characteristic of heterogeneous adsorbents and is observed for the most promising MSUCOF materials in Fig. \ref{fig:qst_analysis} b. MSUCOF-4-FeCp maintains a constant negative slope across a wide range of uptakes/loading, which indicates that it maintains a distribution of binding sites rather than uniform interactions. In contrast, MSUCOF-4 and IRCOF-102 are flat, suggesting that \ce{H2} is mainly bulk filling within the pores. MSUCOF-1-Pt\ce{Cl2} is an interesting case where beyond ~15 g \ce{H2} L$^{-1}$ it quickly flattens as it approaches saturation.

%\onecolumngrid 
% \small

\begin{table}[!htbp]
\small
\centering
\setlength{\tabcolsep}{1pt} % default is 6pt, adjust as needed
\caption{Energy distribution statistics for hydrogen adsorption derived from GCMC simulations at 298 K. $N$ denotes the number of resolved peaks.}
\label{tab:energy_stats}
\begin{tabular}{lccccc}
%\begin{tabular}{l c p{1.1cm} p{0.8cm} p{0.6cm} c}
\hline
\textbf{Material} & \textbf{P} & $\boldsymbol{\langle E \rangle}$ & $\boldsymbol{Q_{\textbf{st}}}$ & $\boldsymbol{\sigma}$ & \textbf{Peaks} \\
\scriptsize & \scriptsize bar & \scriptsize kJ$\cdot$mol$^{-1}$ & \scriptsize kJ$\cdot$mol$^{-1}$ & \scriptsize kJ$\cdot$mol$^{-1}$ & $N$\\
\hline
\multicolumn{6}{l}{\textit{Low pressure}} \\
MSUCOF-1-PtCl$_2$ & 1 & $-17.08$ & 17.08 & 5.18 & 2 \\
MSUCOF-3-CoCl & 1 & $-13.96$ & 13.96 & 2.89 & 1 \\
MSUCOF-3-MnCl & 1 & $-13.68$ & 13.68 & 3.02 & 1 \\
MSUCOF-3-NiCl & 1 & $-14.89$ & 14.89 & 3.36 & 1 \\
MSUCOF-3-PdCl & 1 & $-10.81$ & 10.81 & 2.74 & 1 \\
MSUCOF-3-PtCl & 1 & $-24.53$ & 24.53 & 4.81 & 1 \\
MSUCOF-4-FeCp & 1 & $-17.51$ & 17.51 & 4.79 & 1 \\
\hline
\multicolumn{6}{l}{\textit{Intermediate pressure}} \\
MSUCOF-1-PtCl$_2$ & 17.1 & $-12.97$ & 12.97 & 6.68 & 3 \\
MSUCOF-3-CoCl & 17.1 & $-13.25$ & 13.25 & 3.28 & 2 \\
MSUCOF-3-MnCl & 17.1 & $-12.95$ & 12.95 & 3.39 & 1 \\
MSUCOF-3-NiCl & 17.1 & $-13.83$ & 13.83 & 3.78 & 1 \\
MSUCOF-3-PdCl & 17.1 & $-10.52$ & 10.52 & 2.91 & 1 \\
MSUCOF-3-PtCl & 17.1 & $-22.16$ & 22.16 & 5.58 & 1 \\
MSUCOF-4-FeCp & 17.1 & $-14.99$ & 14.99 & 5.20 & 1 \\
\hline
\multicolumn{6}{l}{\textit{High pressure}} \\
MSUCOF-1-PtCl$_2$ & 700 & $-6.59$ & 6.59 & 6.15 & 1 \\
MSUCOF-3-CoCl & 700 & $-11.62$ & 11.62 & 3.98 & 2 \\
MSUCOF-3-MnCl & 700 & $-11.13$ & 11.13 & 4.13 & 2 \\
MSUCOF-3-NiCl & 700 & $-11.60$ & 11.60 & 4.63 & 2 \\
MSUCOF-3-PdCl & 700 & $-9.36$ & 9.36 & 3.55 & 2 \\
MSUCOF-3-PtCl & 700 & $-19.49$ & 19.49 & 6.16 & 2 \\
MSUCOF-4-FeCp & 700 & $-12.32$ & 12.32 & 5.52 & 2 \\
\hline
\end{tabular}
\end{table}

%\twocolumngrid 
\normalsize

A summary breakdown of the values and statistics in Fig. \ref{fig:energy_dist} is shown in Table \ref{tab:energy_stats} (a more detailed breakdown can be found in Table S1). The dual-peak behavior at high pressures supports the idea of a multi-binding site mechanism, which is key to the exceptional storage capacity of MSUCOF-4-FeCp. Integration of individual peaks reveals that the majority (83\%) of adsorbed \ce{H2} still interact with Fe centers at 700 bar, while the remaining fraction (17\%) engages with weaker framework sites. Hydrogen density fields visualizing the spatial distribution of adsorbed \ce{H2} at 1, 10, and 30 bar confirm that ferrocene sites serve as primary adsorption centers (Fig. \ref{fig:H2Den}). This strong preference for metal coordination demonstrates the critical role of ferrocene centers in achieving high storage densities. In the case of MSUCOF-1-Pt\ce{Cl2}, up to three peaks are observed: one, primary binding site; two, secondary or shielded binding sites; and three, a bulk filled interaction near 0 kJ$\cdot$mol$^{-1}$ as the previous two saturate. 

\subsection{Practical considerations}

Beyond the cost advantages mentioned earlier (Section \ref{sec:cost}), ferrocene functionalization brings additional practical benefits for large-scale implementation. Iron is among the most abundant elements in the Earth's crust, with global reserves exceeding 800 billion metric tonnes of crude ore, while platinum-group metals are scarce and geographically concentrated, with production dominated by South Africa and Russia.\cite{USGS2025MCS} This abundance makes iron-based systems less susceptible to supply chain disruptions and price volatility that affect precious metal-dependent technologies.

MSUCOF-4-FeCp outperforms both pristine COFs and other metal-functionalized variants, while offering significant economic and strategic benefits. The gravimetric uptake of 18.0 wt\% significantly exceeds the DOE target of 6.5 wt\%, while the volumetric uptake of 72.6 g \ce{H2} L$^{-1}$ exceeds the target of 50 g \ce{H2} L$^{-1}$. More importantly, balanced performance metrics ensure high deliverable capacity under practical operating conditions.

The stable and well-defined nature of ferrocene, along with its demonstrated thermal stability up to $400^\circ \text{C}$, suggests excellent cycling stability for pressure swing adsorption applications. It should be noted that these computational results represent idealized conditions: perfect crystallinity without defects, vacancies, or residual solvent molecules; experimental performance may deviate accordingly. Thus, experimental synthesis, optimization, and long-term performance validation are essential for future work.

\section*{Conclusions}

We have demonstrated through comprehensive first-principles multiscale computational simulations that ferrocene-functionalized covalent organic frameworks represent a promising new class of hydrogen storage materials. MSUCOF-4-FeCp achieves exceptional performance metrics with 18.0 wt\% gravimetric and 72.6 g \ce{H2} L$^{-1}$ volumetric hydrogen uptake at 298 K and 700 bar, substantially exceeding DOE targets. More critically, the working capacities of 12.2 wt\% (WC$_g$) and 52.2 g \ce{H2} L$^{-1}$ (WC$_v$) demonstrate that MSUCOF-4-FeCp is, to our knowledge, the only porous framework material to simultaneously satisfy both DOE ultimate targets under practical operating conditions.

Energy distribution analysis from GCMC simulations reveals that iron coordination, the cyclopentadienyl ring, and tritopic region compounded interactions provide multiple hydrogen binding sites while maintaining moderate binding energies. This approach offers significant economic advantages over precious metal alternatives while achieving exceptional performance. The stable and well-defined nature of ferrocene suggests excellent potential for cycling stability and long-term performance.

Ferrocene incorporation narrows the band gap from 4.10 eV to 3.02 eV (410 nm), with the metal-to-ligand charge transfer character creating discrete, localized redox-active sites throughout the framework. These molecular catalytic centers, combined with the porous architecture, suggest applications in photocatalysis and electrocatalysis beyond hydrogen storage.

This work establishes ferrocene functionalization as a viable strategy for enhancing the hydrogen storage performance in COFs and provides a foundation for future experimental validation. The computational framework developed here can be extended to explore other metallocene systems and optimize material properties for specific applications.

\section*{Supplementary Information}
Supplementary Information is available: Surface area and pore volume calculations, force field fitting details, excess uptake isotherms, hydrogen density fields, energy distribution peak deconvolution parameters, and optimized crystal structure geometries.

\section*{Acknowledgements}
This work was partially supported by startup funds from Michigan State University (to J.L.M.-C. and M.D.). 
M.D. was supported by MSU's College of Engineering Graduate Office Fellowship and Dissertation Completion Fellowship.  
M.D. thanks Daniel Maldonado-Lopez for helpful discussions. This work was supported in part through computational resources and services provided by the Institute for Cyber-Enabled Research at Michigan State University. 

\section*{Author contributions}
M.D. performed the DFT calculations, validated the force field parameters, performed the GCMC simulations, analyzed the data, created all figures, and wrote the manuscript. 
J.L.M.-C. conceived the material design, developed the initial methodology and results, supervised the project, co-wrote and revised the manuscript. 
Both authors discussed the results and approved the final version.

\section*{Competing interests}
A patent application related to this work has been filed and is currently pending.

\section*{Data availability}
Data and codes supporting the findings of this study are available from the corresponding author on a reasonable request. 

\bibliographystyle{naturemag}
\bibliography{bibliography}

\clearpage
\newpage

\renewcommand{\thepage}{S\arabic{page}}
\renewcommand{\thesection}{S\arabic{section}}
\renewcommand{\thetable}{S\arabic{table}}
\renewcommand{\thefigure}{S\arabic{figure}}
\renewcommand{\theequation}{S\arabic{equation}}

\setcounter{section}{0}
\setcounter{figure}{0}
\setcounter{table}{0}

\renewcommand{\theHfigure}{S\arabic{figure}}
\renewcommand{\theHtable}{S\arabic{table}}
\renewcommand{\theHequation}{S\arabic{equation}}
\renewcommand{\theHsection}{S\arabic{section}}

\onecolumngrid 
\begin{center}
{\LARGE \textbf{Supplementary Information (SI)}}
\end{center}

\vspace{1em}

\noindent\rule{\textwidth}{1.5pt}

\begin{center}
\Large{\textbf{Ferrocene-functionalized covalent organic frameworks exceeding the ultimate hydrogen storage targets: a first-principles multiscale computational study}}

\vspace{1em}

\large
Marcus Djokic$^{a}$ and Jose L. Mendoza-Cortes$^{a,b,*}$

\vspace{1em}

\normalsize
$^{a}$~Department of Chemical Engineering and Materials Science, Michigan State University, East Lansing, MI 48824, USA\\[0.3em]
$^{b}$~Department of Physics and Astronomy, Michigan State University, East Lansing, MI 48824, USA\\[0.5em]
$^{*}$~E-mail: jmendoza@msu.edu

\end{center}

\noindent\rule{\textwidth}{1.5pt}

\vspace{2em}

\tableofcontents

\newpage

\newpage
\section{Surface Area and Pore Volume}\label{sec:SA}
%\addcontentsline{toc}{section}{Surface Area and Pore Volume}

\begin{figure}[htp!]
	%	\vspace{3mm}
	\includegraphics[width=0.80\linewidth]{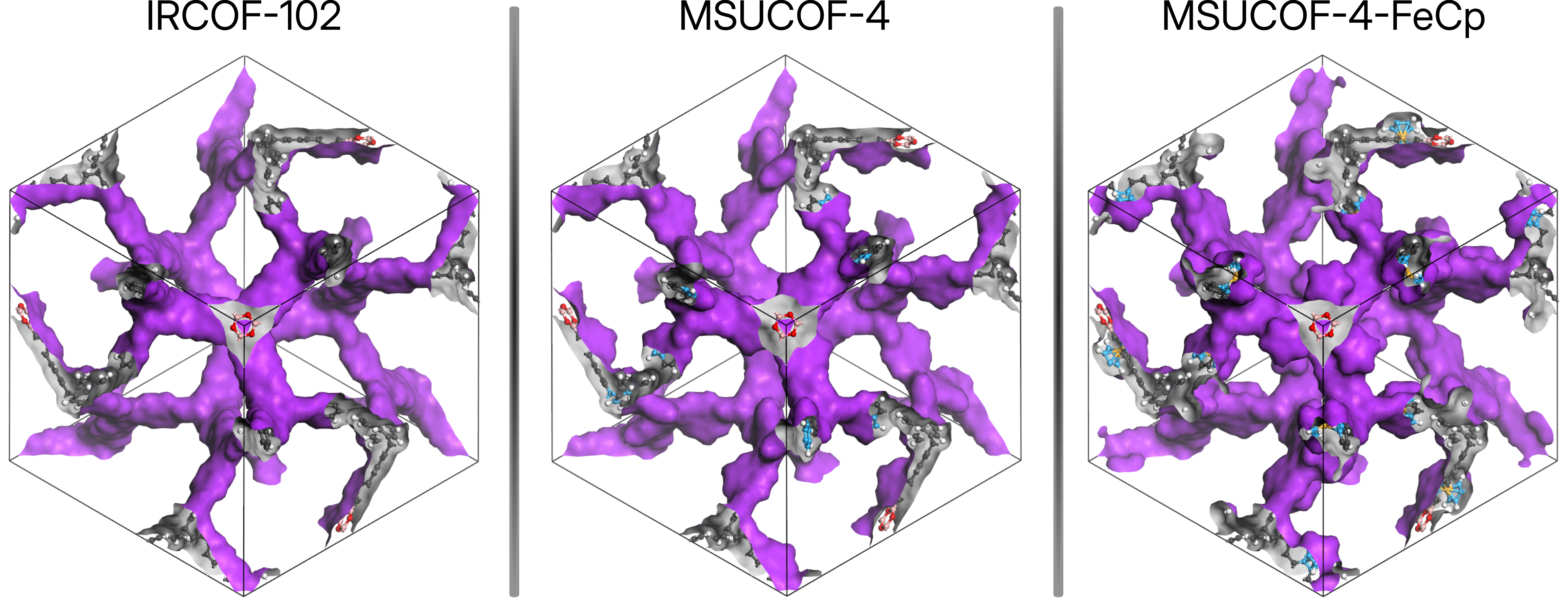}
	%	\vspace{5mm}
	\caption{Van der Waal surface (in purple) projected onto IRCOF-102, MSUCOF-4, and MSUCOF-4-FeCp. This surface is used for pore volume and surface area calculations.}%Van der Waal surface (in blue) projected onto a) MSUCOF-1 and b) metalated example MSUCOF-1-Pd\ce{Cl2}. This surface is used for pore volume and surface area calculations.}
	\label{fig:vdw}
	%	\vspace{-10mm}  
\end{figure}

Surface areas and pore volumes were calculated using the Atom Volumes \& Surfaces 
tool in Materials Studio.\cite{biovia_materials_2022_updated} The van der Waals (VdW) 
surface is constructed by rolling a spherical probe across the framework atoms, 
where the probe radius is defined by the adsorbate of interest. For hydrogen 
storage applications, a probe radius of 1.445 \AA\ was used, corresponding 
to half the kinetic diameter of \ce{H2} (2.89 \AA). The resulting VdW surface 
(Fig.~\ref{fig:vdw}) delineates the accessible pore volume and the surface area 
available for hydrogen adsorption. This approach provides a consistent basis 
for comparing geometric properties across the IRCOF-102, MSUCOF-4, and 
MSUCOF-4-FeCp frameworks.

\newpage
\section{Force Field Fitting}
%\addcontentsline{toc}{section}{Force Field Fitting}

Ferrocene adopts two limiting conformations: eclipsed and staggered, with the latter slightly favored in isolation. However, within the 
MSUCOF-4-FeCp framework, the spatial constraints imposed by the neighboring struts 
stabilize the eclipsed conformation. Therefore, both conformers were included 
in the force field training set to ensure transferability.

\begin{figure}[htp!]
	%	\vspace{3mm}
	\includegraphics[width=0.8\linewidth]{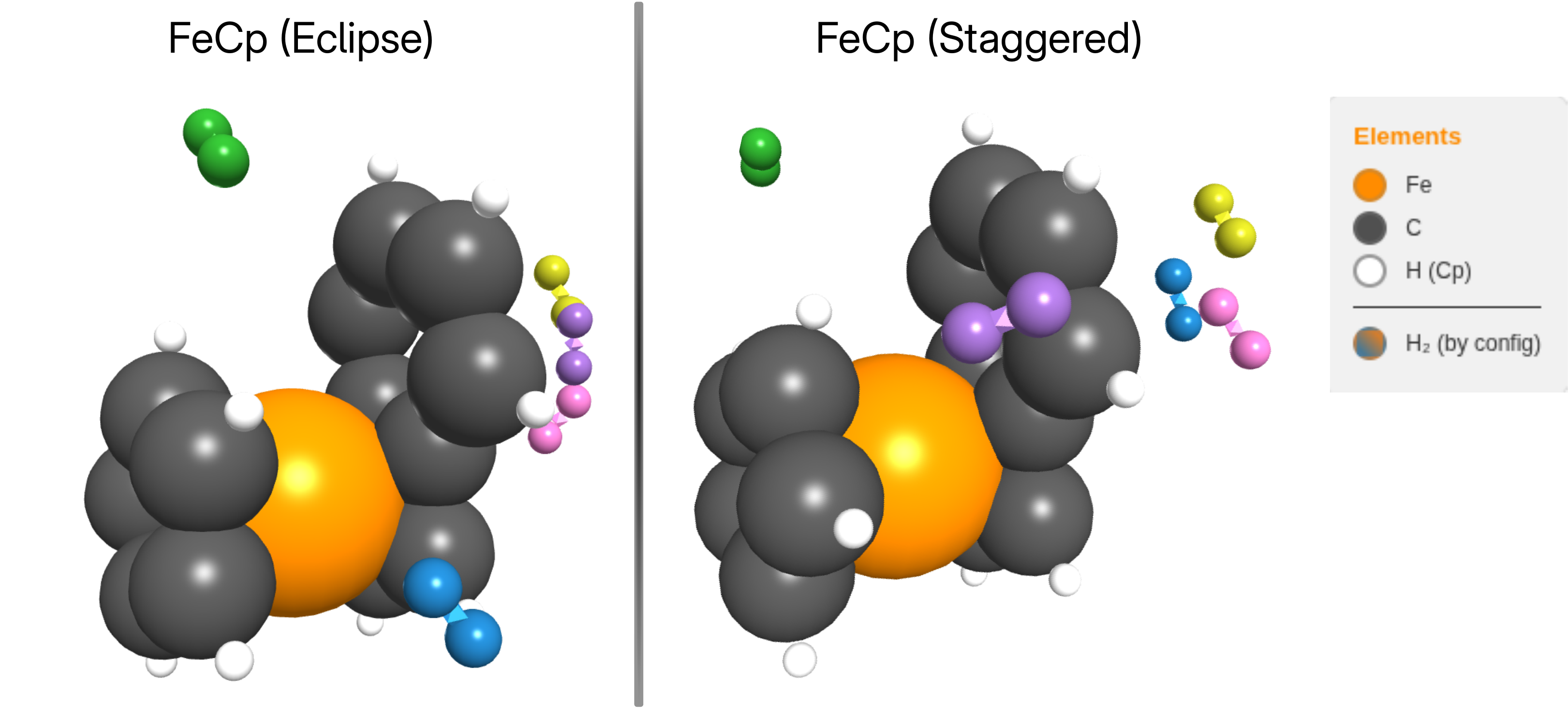}
	%	\vspace{5mm}
	\caption{We show example configurations that were used in the training of the QM-based force field fitting. Only the top 5 configuration are shown for display purposes. Note the hydrogen molecules are colored based on the configuration displayed.}
	\label{fig:configurations} 
	%	\vspace{-10mm}  
\end{figure}

The initial \ce{H2} binding configurations were identified using the Adsorption 
Locator tool in Materials Studio \cite{biovia_materials_2022_updated} which uses 
Monte Carlo sampling to locate energetically favorable adsorption sites. A 
random selection of these configurations was extracted, spanning various \ce{H2} distances 
and orientations relative to the ferrocene center, for quantum 
mechanical refinement (Fig.~\ref{fig:configurations}).

Fragment calculations were performed following the methodology described in 
Section \ref{sec:fragDFT} % will need to reformat to match journal 
of the main text. Vibrational frequency analysis confirmed the 
absence of imaginary frequencies and provided thermodynamic corrections, 
including zero-point energy (ZPE) and vibrational enthalpy ($\Delta H^{o}_{vib}$). 
Binding enthalpies were calculated as follows:

\begin{equation} \label{eq:bind}
\centering
\Delta H^{o}_{bind} = \Delta H^{o}_{Linker + \ce{H2}} - \Delta H^{o}_{Linker} - \Delta H^{o}_{\ce{H2}}
\end{equation}

\noindent where $\Delta H^{o}$ represents the sum of the electronic energy, ZPE, 
and the correction of the vibrational enthalpy. The resulting QM binding energies were 
used to parameterize the Morse potentials using least-squares fitting in GULP, \cite{GULP} 
achieving mean absolute errors below 0.5 kJ mol$^{-1}$ (Fig. \ref{fig:ff_validation}).% Main Manuscript reference

\clearpage
\newpage
\section{Isotherm Plots}
%\addcontentsline{toc}{section}{Isotherm Plots}

\begin{figure}[!htb]
\centering
\includegraphics[width=\textwidth]{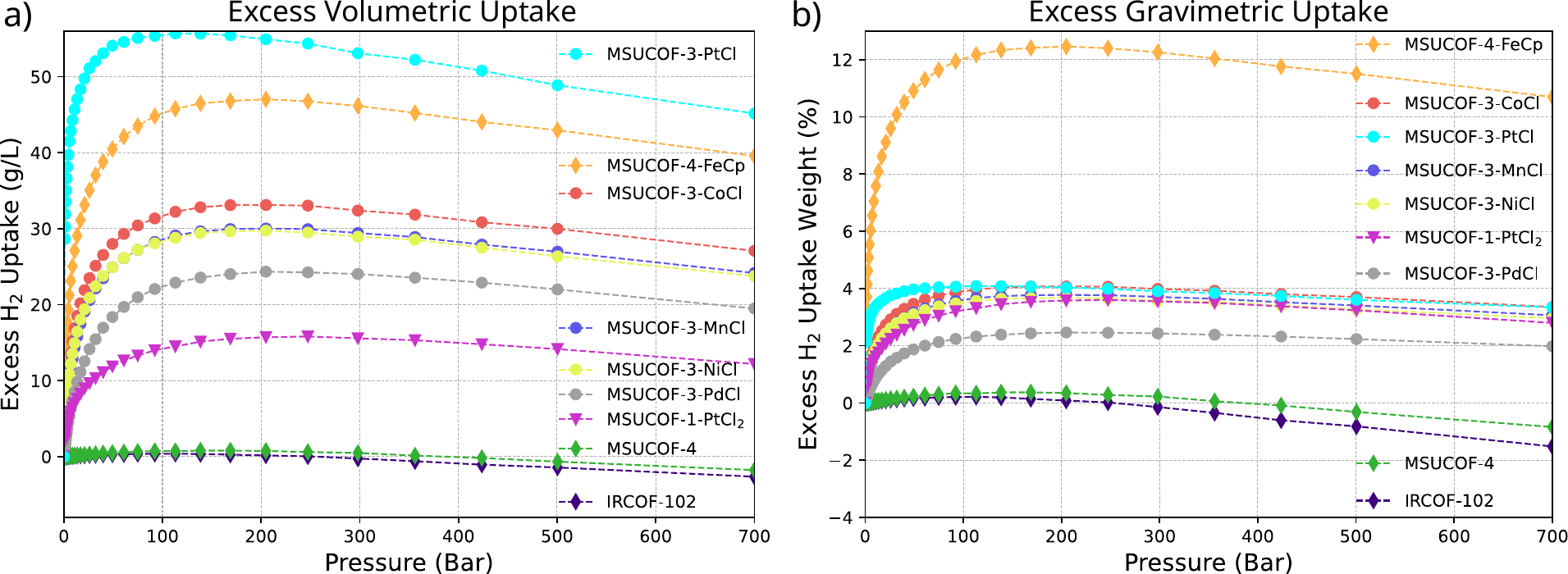}
\caption{Comparison of excess (a) volumetric and (b) gravimetric uptake at 298 K for the MSUCOF-4 family against top-performing MSUCOF variants. Excess 
uptake subtracts the bulk \ce{H2} density corresponding to the pore volume, 
whereas net uptake (main text) applies a stricter correction using the full 
unit cell volume.}
\label{fig:excess_isotherm}
\end{figure}

For completeness, we present the excess uptake isotherms (Fig.~\ref{fig:excess_isotherm}) 
to complement the total and net uptake data in the main text. The distinction between 
these metrics deserves clarification.

\textbf{Net uptake} subtracts the amount of \ce{H2} that would occupy the entire volume of the unit 
cell if no framework was present (i.e., bulk gas under equivalent conditions). 
This metric directly quantifies the storage advantage of incorporating a porous material 
versus an empty tank of identical dimensions, making it the most rigorous computational 
benchmark. By contrast, excess uptake subtracts only the bulk \ce{H2} that 
would occupy the accessible pore volume, excluding the volume displaced by the framework 
atoms.

From an experimental standpoint, excess uptake more closely reflects what is measured 
in typical adsorption experiments, as isolating the contribution of the full unit cell 
volume is not straightforward. However, computationally derived excess uptake is 
sensitive to how the pore volume is defined; specifically the choice of probe radius 
and van der Waals surface construction. Net uptake avoids 
this ambiguity by using the crystallographically defined unit cell volume. Hydrogen density values at each pressure and temperature were obtained from the 
NIST Chemistry WebBook.\cite{linstrom_nist_2001}

\newpage
\section{Hydrogen Density In MSUCOFs}
%\addcontentsline{toc}{section}{Hydrogen Density In MSUCOFs}

\begin{figure}[htp!]
	%	\vspace{3mm}
	\includegraphics[width=1\linewidth]{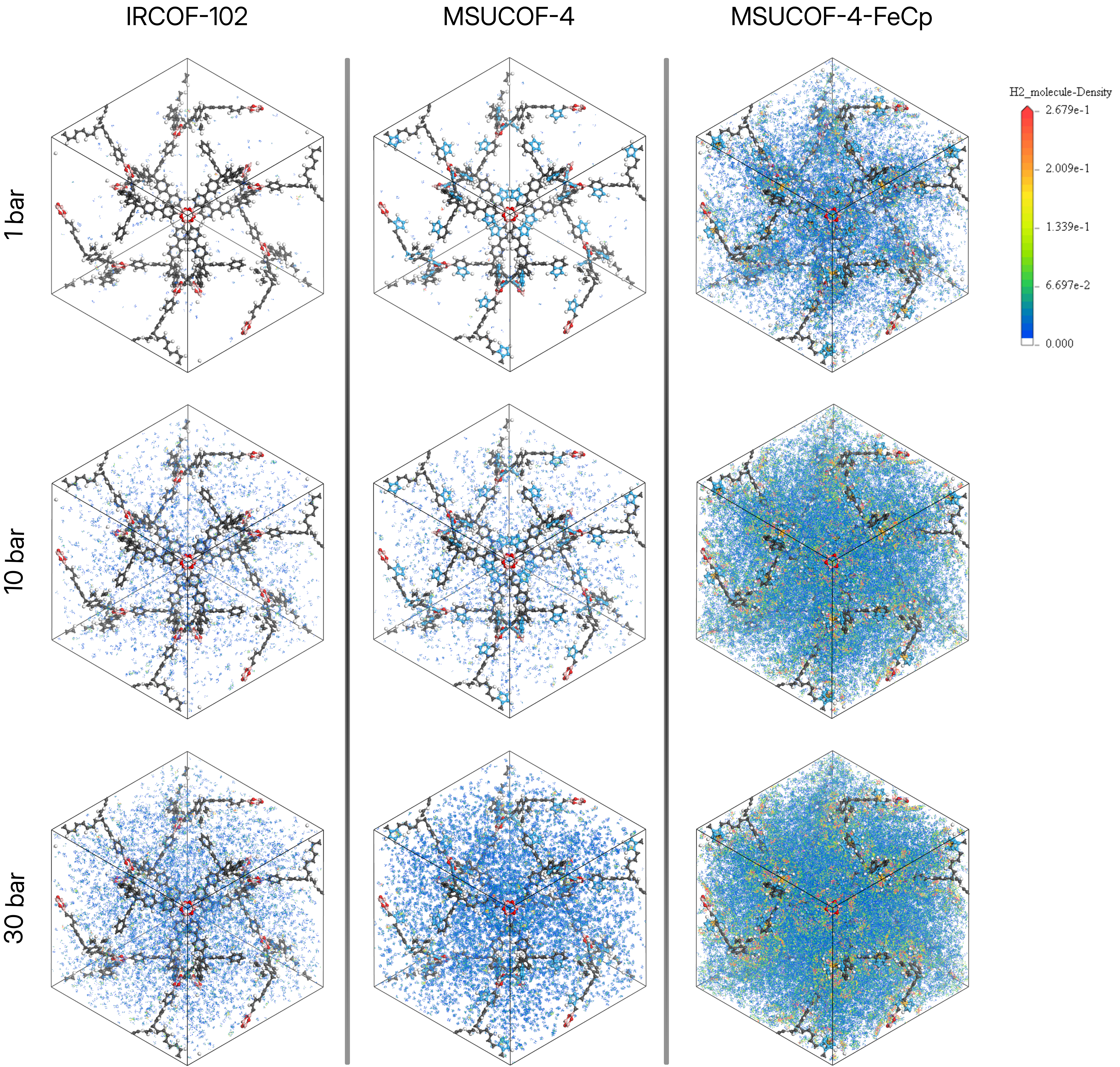}
	%	\vspace{5mm}
	\caption{Hydrogen density field displayed in IRCOF-102, MSUCOF-4, and MSUCOF-4-FeCp. The hydrogen density field is colored according to the rainbow legend. The fields are shown at various pressures (1, 10, and 30 bar) to demonstrate how the fields change upon further loading.}%Hydrogen density field displayed in the pristine and metallated example of MSUCOF-1 and MSUCOF-3. MSUCOFs are displayed as a purple space-filling model while the hydrogen density field is colored according to the rainbow legend. The fields are shown at various pressures (1, 10, and 30 bar) to demonstrate how the fields change upon further loading.}
	\label{fig:H2Den}
	%	\vspace{-10mm}  
\end{figure}

Hydrogen density fields provide spatial insight into adsorption behavior that 
complements the isotherm data. These fields (Fig.~\ref{fig:H2Den}) were generated 
using the Sorption module in Materials Studio,\cite{biovia_materials_2022_updated} where 
the local \ce{H2} density at each point in the grid is calculated by averaging the 
count of sorbates over the associated volume element during the GCMC sampling.

At low pressure (1 bar), minimal \ce{H2} loading is observed in all three 
frameworks. IRCOF-102 and MSUCOF-4 show negligible density throughout the pores, 
while MSUCOF-4-FeCp already exhibits a low but discernible density (blue regions, 
$<5 \times 10^{-2}$ molecules \AA$^{-3}$) localized near the ferrocene centers 
and tritopic binding regions. This early-stage preferential adsorption reflects 
the enhanced binding affinity of the Fe sites.

At an intermediate pressure (10 bar), the contrast becomes pronounced. The pristine 
IRCOF-102 and MSUCOF-4 frameworks show only modest increases in \ce{H2} density, 
with adsorption still diffuse throughout the pore volume. In contrast, MSUCOF-4-FeCp 
displays significant saturation of the ferrocene coordination sites, evidenced by 
elevated densities (green--red regions, $>5 \times 10^{-2}$ molecules \AA$^{-3}$) 
concentrated around the FeCp moieties, while the remainder of the pores transition 
to a uniformly low density (blue). This spatial heterogeneity confirms that the ferrocene 
sites serve as primary adsorption centers.

At 30 bar, the pristine IRCOF-102 and MSUCOF-4 begin to show appreciable loading, 
with density accumulating near the framework surfaces and tritopic regions. 
However, MSUCOF-4-FeCp exhibits substantially higher saturation throughout, 
with the ferrocene sites approaching capacity. Relative hydrogen densities 
follow a clear hierarchy: MSUCOF-4-FeCp $\gg$ MSUCOF-4 $>$ IRCOF-102, consistent 
with isotherm data and confirming that ferrocene functionalization dramatically 
improves \ce{H2} uptake at moderate pressures.

\clearpage
\newpage
\section{Energy Distribution Peak Deconvolution}
%\addcontentsline{toc}{section}{Energy Distribution Peak Deconvolution}

Table~\ref{tab:peak_deconvolution} provides an extended breakdown of the energy distribution analysis summarized in the main text (Table \ref{tab:energy_stats}). 
While the main manuscript reports ensemble-averaged quantities ($\langle E \rangle$, $Q_{\text{st}}$, $\sigma$), this section details the individual parameters for each deconvoluted peak. These parameters are obtained by fitting asymmetric Gaussian functions to the interaction energy distributions.

For each peak $i$, we report the peak center energy ($\mu_i$), the fraction of adsorbed \ce{H2} molecules associated with that binding mode ($f_i$), the site-specific isosteric heat ($Q_{\text{st},i} = -\mu_i$), and the full width at half maximum (FWHM). The population fraction $f_i$ is calculated via numerical integration of the raw distribution within the specified energy boundaries. Peaks are classified as ``Main'' when they correspond to primary binding sites (e.g., metal centers, tritopic regions) or contribute significantly to the total uptake ($f_i > 0.30$). ``Secondary'' peaks reflect lower-energy configurations, including shielded sites, \ce{H2}--\ce{H2} interactions, or bulk-like pore filling at high pressures.

The emergence of multiple peaks at elevated pressures reflects the sequential saturation of binding sites. Primary sites (strongest affinity) generally saturate first, followed by the population of progressively weaker secondary sites. This detailed breakdown enables a quantitative assessment of how distinct binding environments contribute to the overall macroscopic adsorption isotherm.

\begin{table}[htbp]
\small
\centering
\setlength{\tabcolsep}{4pt}
\caption{Detailed peak deconvolution parameters from GCMC simulations at 298 K. $\mu_i$: peak center energy; $f_i$: population fraction; $Q_{\text{st},i}$: site-specific isosteric heat; FWHM: full width at half maximum; Boundary: $[E_{min}, E_{max}]$ used for integration.}
\label{tab:peak_deconvolution}
\begin{tabular}{lcclccccc}
\hline
\textbf{Material} & \textbf{P} & \textbf{Peak} & \textbf{Type} & $\boldsymbol{\mu_i}$ & $\boldsymbol{f_i}$ & $\boldsymbol{Q_{\textbf{st},i}}$ & \textbf{FWHM} & \textbf{Boundary} \\
 & (bar) & & & \scriptsize(kJ/mol) & & \scriptsize(kJ/mol) & \scriptsize(kJ/mol) & \scriptsize(kJ/mol) \\
\hline
\multicolumn{9}{l}{\textit{Low pressure (1 bar)}} \\
MSUCOF-1-PtCl$_2$ & 1.0 & 1 & Main & $-21.13$ & 0.86 & 22.15 & 7.09 & [$-25.31, 4.81$] \\
MSUCOF-1-PtCl$_2$ & 1.0 & 2 & Secondary & $-0.63$ & 0.14 & 0.21 & 12.62 & [$-2.72, 4.81$] \\
MSUCOF-3-CoCl & 1.0 & 1 & Main & $-15.69$ & 0.97 & 16.29 & 5.76 & [$-19.87, 6.07$] \\
MSUCOF-3-MnCl & 1.0 & 1 & Main & $-15.27$ & 0.97 & 16.03 & 5.92 & [$-19.46, 5.65$] \\
MSUCOF-3-NiCl & 1.0 & 1 & Main & $-16.95$ & 0.97 & 17.22 & 6.52 & [$-22.38, 8.16$] \\
MSUCOF-3-PdCl & 1.0 & 1 & Main & $-12.76$ & 0.99 & 12.81 & 6.02 & [$-16.95, 9.41$] \\
MSUCOF-3-PtCl & 1.0 & 1 & Main & $-27.82$ & 1.00 & 27.64 & 11.22 & [$-37.03, 0.21$] \\
MSUCOF-4-FeCp & 1.0 & 1 & Main & $-19.46$ & 1.00 & 18.10 & 11.45 & [$-29.50, 2.72$] \\
\hline
\multicolumn{9}{l}{\textit{Intermediate pressure (17.1 bar)}} \\
MSUCOF-1-PtCl$_2$ & 17.1 & 1 & Main & $-16.95$ & 0.81 & 20.06 & 11.95 & [$-25.73, 8.58$] \\
MSUCOF-1-PtCl$_2$ & 17.1 & 2 & Secondary & $-4.39$ & 0.13 & 2.70 & 4.93 & [$-7.32, -3.14$] \\
MSUCOF-1-PtCl$_2$ & 17.1 & 3 & Secondary & $-0.63$ & 0.06 & 0.29 & 1.15 & [$-3.14, 8.58$] \\
MSUCOF-3-CoCl & 17.1 & 1 & Main & $-15.69$ & 0.76 & 15.73 & 5.21 & [$-20.29, 10.25$] \\
MSUCOF-3-CoCl & 17.1 & 2 & Secondary & $-9.83$ & 0.24 & 8.66 & 5.21 & [$-10.25, 7.32$] \\
MSUCOF-3-MnCl & 17.1 & 1 & Main & $-15.27$ & 0.99 & 15.70 & 7.28 & [$-19.87, 9.00$] \\
MSUCOF-3-NiCl & 17.1 & 1 & Main & $-16.11$ & 0.98 & 16.55 & 8.00 & [$-23.22, 8.16$] \\
MSUCOF-3-PdCl & 17.1 & 1 & Main & $-12.34$ & 1.00 & 12.78 & 6.66 & [$-17.36, 10.25$] \\
MSUCOF-3-PtCl & 17.1 & 1 & Main & $-23.64$ & 1.00 & 24.86 & 13.54 & [$-37.45, 5.23$] \\
MSUCOF-4-FeCp & 17.1 & 1 & Main & $-15.69$ & 1.00 & 15.16 & 13.23 & [$-30.33, 9.00$] \\
\hline
\multicolumn{9}{l}{\textit{High pressure (700 bar)}} \\
MSUCOF-1-PtCl$_2$ & 700.0 & 1 & Main & $-1.46$ & 0.78 & 1.14 & 6.68 & [$-25.73, 18.62$] \\
MSUCOF-3-CoCl & 700.0 & 1 & Main & $-14.85$ & 0.90 & 14.90 & 8.97 & [$-21.13, 11.92$] \\
MSUCOF-3-CoCl & 700.0 & 2 & Secondary & $-8.16$ & 0.10 & 8.06 & 3.05 & [$-9.41, 10.25$] \\
MSUCOF-3-MnCl & 700.0 & 1 & Main & $-14.02$ & 0.92 & 14.35 & 9.17 & [$-20.71, 13.60$] \\
MSUCOF-3-MnCl & 700.0 & 2 & Secondary & $-6.49$ & 0.08 & 6.51 & 2.73 & [$-7.74, 11.51$] \\
MSUCOF-3-NiCl & 700.0 & 1 & Main & $-14.02$ & 0.58 & 14.54 & 6.68 & [$-23.64, 14.85$] \\
MSUCOF-3-NiCl & 700.0 & 2 & Main & $-5.65$ & 0.42 & 5.89 & 8.64 & [$-6.90, 11.92$] \\
MSUCOF-3-PdCl & 700.0 & 1 & Main & $-11.92$ & 0.92 & 12.09 & 8.17 & [$-18.20, 13.18$] \\
MSUCOF-3-PdCl & 700.0 & 2 & Secondary & $-6.07$ & 0.08 & 5.79 & 2.69 & [$-6.90, 11.51$] \\
MSUCOF-3-PtCl & 700.0 & 1 & Main & $-21.97$ & 0.87 & 22.74 & 13.15 & [$-37.87, 11.51$] \\
MSUCOF-3-PtCl & 700.0 & 2 & Secondary & $-12.76$ & 0.13 & 12.50 & 5.38 & [$-15.27, 7.74$] \\
MSUCOF-4-FeCp & 700.0 & 1 & Main & $-13.18$ & 0.83 & 13.08 & 13.19 & [$-30.75, 13.60$] \\
MSUCOF-4-FeCp & 700.0 & 2 & Secondary & $-7.74$ & 0.17 & 7.80 & 3.34 & [$-11.51, 10.67$] \\
\hline
\end{tabular}
\end{table}

\newpage
\section{Optimized Geometries} \label{sec:cif}
%\addcontentsline{toc}{section}{Optimized Geometries}
Here we report the optimized structures obtained at the hybrid-DFT level of theory (HSE06-D3). 

\footnotesize

\subsection{IRCOF-102}
\begin{verbatim}
data_IRCOF102
_cell_length_a    42.955631
_cell_length_b    42.955631
_cell_length_c    42.955631
_cell_angle_alpha 90.000000
_cell_angle_beta  90.000000
_cell_angle_gamma 90.000000
_space_group_name_H-M_alt       'I -4 3 d'
_space_group_IT_number          220
loop_
_atom_site_label
_atom_site_type_symbol
_atom_site_fract_x
_atom_site_fract_y
_atom_site_fract_z
O1       O      0.210053    -0.195275     0.260431
H1       H      0.049110    -0.208302     0.118102
H2       H      0.109203    -0.280245     0.222018
H3       H      0.095876    -0.203198     0.150390
H4       H      0.014119    -0.278596     0.179640
H5       H      0.153924    -0.272298     0.257309
H6       H      0.060078    -0.272257     0.213170
C1       C      0.028534    -0.244580     0.145792
C2       C      0.032412    -0.262350     0.172908
C3       C      0.058413    -0.258813     0.191739
C4       C      0.077914    -0.219722     0.157247
C5       C      0.081877    -0.237391     0.184295
C6       C      0.150987    -0.200698     0.224463
C7       C      0.051708    -0.223026     0.138554
H7       H      0.163010    -0.178361     0.224930
C8       C      0.120450    -0.257547     0.223164
C9       C      0.161779    -0.224795     0.243581
H8       H      0.116331    -0.185073     0.192053
C10      C      0.125032    -0.204616     0.205657
C11      C      0.109341    -0.233211     0.204465
C12      C      0.145846    -0.253224     0.242656
B1       B      0.189921    -0.219831     0.265204
C13      C      0.250000     0.375000    -0.000000

\end{verbatim}

\clearpage
\newpage

\subsection{MSUCOF-4}
\begin{verbatim}
data_MSUCOF-4
_cell_length_a    42.974269
_cell_length_b    42.974269
_cell_length_c    42.974269
_cell_angle_alpha 90.000000
_cell_angle_beta  90.000000
_cell_angle_gamma 90.000000
_space_group_name_H-M_alt       'I -4 3 d'
_space_group_IT_number          220
loop_
_atom_site_label
_atom_site_type_symbol
_atom_site_fract_x
_atom_site_fract_y
_atom_site_fract_z
H1       H     -0.158135     0.237213     0.179336
H2       H     -0.126859     0.206083     0.135292
C1       C     -0.173131     0.198626     0.113882
C2       C     -0.168023     0.225856     0.159030
C3       C     -0.151879     0.209499     0.136080
H3       H     -0.167510     0.186439     0.092456
O1       O     -0.199299     0.256263     0.215188
H4       H     -0.209918     0.116670     0.050006
H5       H     -0.282069     0.214545     0.114075
H6       H     -0.277727     0.180318     0.013834
H7       H     -0.271326     0.213562     0.060219
H8       H     -0.275895     0.247437     0.160607
H9       H     -0.205337     0.148686     0.097424
C4       C     -0.261937     0.173155     0.032343
C5       C     -0.237545     0.183681     0.082152
C6       C     -0.244714     0.145663     0.028660
C7       C     -0.223996     0.137589     0.052318
C8       C     -0.226409     0.240157     0.165058
C9       C     -0.204132     0.207996     0.123301
C10      C     -0.258430     0.191802     0.058482
C11      C     -0.233334     0.203390     0.109804
C12      C     -0.220766     0.156008     0.078699
C13      C     -0.255529     0.236332     0.150353
C14      C     -0.200907     0.225605     0.151186
B1       B     -0.223173     0.260696     0.194234
C15      C     -0.258998     0.218213     0.123924
C16      C      0.375000    -0.000000     0.250000
\end{verbatim}

\clearpage
\newpage

\subsection{MSUCOF-4-FeCp}
\begin{verbatim}
data_MSUCOF-4-FeCp
_cell_length_a    43.072089
_cell_length_b    43.072089
_cell_length_c    43.072089
_cell_angle_alpha 90.000000
_cell_angle_beta  90.000000
_cell_angle_gamma 90.000000
_space_group_name_H-M_alt       'I -4 3 d'
_space_group_IT_number          220
loop_
_atom_site_label
_atom_site_type_symbol
_atom_site_fract_x
_atom_site_fract_y
_atom_site_fract_z
H1       H     -0.134982     0.149227     0.179180
H2       H     -0.231777     0.146943     0.148759
H3       H     -0.227735     0.185645     0.197590
H4       H     -0.174239     0.123904     0.137467
C1       C     -0.159911     0.152672     0.178400
C2       C     -0.177526     0.172131     0.198639
C3       C     -0.180697     0.139432     0.156223
Fe1      Fe    -0.181730     0.186482     0.154303
H5       H     -0.155773     0.241797     0.175299
C4       C     -0.211072     0.150953     0.162574
H6       H     -0.124740     0.208608     0.132020
C5       C     -0.171328     0.200853     0.110626
C6       C     -0.165725     0.230594     0.155072
C7       C     -0.149520     0.213107     0.132201
H7       H     -0.165942     0.186504     0.090680
H8       H     -0.168335     0.186405     0.217235
O1       O     -0.197264     0.258616     0.211833
H9       H     -0.208529     0.118346     0.049133
H10      H     -0.280252     0.215410     0.113213
C8       C     -0.209070     0.171176     0.188773
H11      H     -0.278985     0.179179     0.013835
H12      H     -0.271923     0.213677     0.059128
H13      H     -0.273482     0.249930     0.158721
H14      H     -0.202769     0.152028     0.095268
C9       C     -0.262580     0.172844     0.032031
C10      C     -0.236742     0.185275     0.080468
C11      C     -0.244752     0.145810     0.028432
C12      C     -0.223084     0.138868     0.051471
C13      C     -0.224519     0.244384     0.161746
C14      C     -0.201644     0.211641     0.119487
C15      C     -0.258673     0.192194     0.057473
C16      C     -0.232184     0.205807     0.107339
C17      C     -0.219291     0.158094     0.077119
C18      C     -0.253026     0.239326     0.148211
C19      C     -0.198044     0.230386     0.147118
B1       B     -0.221218     0.264045     0.191289
C20      C     -0.256908     0.220033     0.121792
C21      C      0.375000     0.000000     0.250000
\end{verbatim}

%\end{multicols}

\clearpage
\newpage

% \phantomsection
% \addcontentsline{toc}{section}{References}
% \bibliographystyle{unsrt}
% \bibliography{ForArXiv/bibliography}

\end{document}